\documentclass[a4paper,11pt]{article}
\usepackage[dvips]{graphicx}
\usepackage{float}
\usepackage{epsfig}
\usepackage{mathtools}
\usepackage{mathbbol}
\usepackage{amssymb}
\usepackage{amsthm}
\usepackage{soul}
\usepackage{empheq}
\usepackage{latexsym,amsmath,amsfonts,amssymb}
\usepackage[labelfont=bf]{caption}
\usepackage{rotating}
\usepackage[american]{babel}
\usepackage[dvips]{graphicx}
\usepackage[section]{placeins}
\usepackage{bbm}
\usepackage{color}
\usepackage{slashed}
\usepackage{lscape}
\usepackage{bigints}
\usepackage{enumerate}
\usepackage[shortlabels]{enumitem}
\usepackage{subcaption}
\usepackage{tikz}
\usetikzlibrary{decorations.pathreplacing}
\usetikzlibrary{shapes}
\usepackage{soul}
\usepackage{graphicx}
\usepackage{epsfig}
\usepackage{rotating}
\usepackage{amssymb}
\usepackage{dsfont}
\usepackage{psfrag}
\usepackage{amsmath,euscript,array,mathrsfs,amsfonts}
\usepackage{slashed}
\usepackage{array}
\usepackage{youngtab}
\usepackage{color}
\usepackage{bbold}
\usepackage[dvipsnames]{xcolor}
\usepackage[numbers,sort&compress]{natbib}
\usepackage{hyperref}
\hypersetup{
colorlinks = true,
linkcolor = purple,
linktocpage = true,
citecolor = blue,
urlcolor = blue
}

\usepackage{tikz}
\usetikzlibrary{calc}
\usetikzlibrary{decorations.text}
\usetikzlibrary{shapes}
\usetikzlibrary{decorations.pathmorphing}
\usetikzlibrary{decorations.pathreplacing}
\usetikzlibrary{arrows.meta}
\tikzset{
 >={To[length=5pt]}
 }
\usetikzlibrary{shapes, shapes.geometric, shapes.symbols, shapes.arrows, shapes.multipart, shapes.callouts, shapes.misc}
\tikzset{snake it/.style={decorate, decoration=snake}}
\tikzset{7brane/.style={circle, draw=black, fill=black,ultra thick,inner sep=1.5 pt, minimum size=1 pt,}, c/.default={4pt}}
\tikzset{cross/.style={cross out, draw=black,thick, minimum size=2*(#1-\pgflinewidth), inner sep=0pt, outer sep=0pt}, cross/.default={5pt}}
\tikzset{big7brane/.style={circle, draw=black, fill=black,ultra thick,inner sep=2.5 pt, minimum size=1 pt,}, c/.default={4pt}}
\tikzset{u/.style={circle, draw=black, fill=white,inner sep=2 pt, minimum size=2 pt,},f/.style={square, draw=black, fill=white,ultra thick,inner sep=4 pt, minimum size=2 pt,}}
\tikzset{so/.style={circle, draw=black, fill=red,inner sep=2 pt, minimum size=2 pt,},f/.style={square, draw=black, fill=white,ultra thick,inner sep=4 pt, minimum size=2 pt,}}
\tikzset{sp/.style={circle, draw=black, fill=blue,inner sep=2 pt, minimum size=2 pt,},f/.style={square, draw=black, fill=white,ultra thick,inner sep=4 pt, minimum size=2 pt,}}
\tikzset{uf/.style={rectangle, draw=black, fill=white,inner sep=3 pt, minimum size=4 pt,}}
\tikzset{spf/.style={rectangle, draw=black, fill=blue, thick,inner sep=3 pt, minimum size=4 pt, circle, draw=black, fill=blue,thick,inner sep=2 pt, minimum size=2 pt,},f/.style={square, draw=black, fill=white,ultra thick,inner sep=4 pt, minimum size=2 pt,}}
\tikzset{sof/.style={rectangle, draw=black, fill=red, thick,inner sep=3 pt, minimum size=4 pt,}}
\usetikzlibrary{positioning}
\usetikzlibrary{arrows}
\usetikzlibrary{decorations.pathreplacing}
\usetikzlibrary{shapes}
 
\makeatletter\def\l@subsubsection#1#2{}
\makeatother
\renewcommand\theequation{\arabic{section}.\arabic{equation}}
\def\CC{\ensuremath{\mathds C}}
\def\RR{\ensuremath{\mathds R}}
\def\ZZ{\ensuremath{\mathds Z}}

\DeclareMathOperator{\sech}{sech}

\DeclareMathOperator{\csch}{csch}

\newcommand{\be}{\begin{equation}}
\newcommand{\ee}{\end{equation}}
\newcommand{\ba}{\begin{array}}
\newcommand{\ea}{\end{array}}
\newcommand{\eq}[1]{eq.~(\ref{#1})}

\def\im{Invent. Math.}

\def\d{\delta}
\def\f{\phi} 

\def\j{\psi}

 \def\th{\theta} 
\def\r{\rho} 

\def\6{\partial}

\DeclareMathOperator{\str}{str}

\newcommand{\beq}{\begin{equation}}
\newcommand{\eeq}{\end{equation}}
\newcommand{\bea}{\begin{eqnarray}}
\newcommand{\eea}{\end{eqnarray}}

\newcommand{\beqs}{\begin{eqnarray}}
\newcommand{\eeqs}{\end{eqnarray}}
\newcommand{\bal}{\begin{aligned}}
\newcommand{\eal}{\end{aligned}}
\makeatletter
\newcommand\setItemnumber[1]{\setcounter{enum\romannumeral\@enumdepth}{\numexpr#1-1\relax}}
\makeatother
\def\del{{\partial}}

\def\Poincaré{{Poincar\'e }}

\def\lbldef#1#2{\expandafter\gdef\csname #1\endcsname {#2}}

\newcommand{\ber}{\begin{eqnarray}}
\newcommand{\eer}{\end{eqnarray}}
\newcommand{\beqar}{\begin{eqnarray}}
\newcommand{\eeqar}{\end{eqnarray}}

\newcommand{\dsl}
{\kern.06em\hbox{\raise.15ex\hbox{$/$}\kern-.56em\hbox{$\partial$}}}

\newcommand{\eeqarr}{\end{eqnarray}}
\def\del{{\delta^{\hbox{\sevenrm B}}}}

\def\im{{\hbox{\rm Im}}}

\def\ie{{\em i.e.}}
\def\ie{\hbox{\it i.e.}}

\def\CC{{\mathchoice
{\rm C\mkern-8mu\vrule height1.45ex depth-.05ex
width.05em\mkern9mu\kern-.05em}
{\rm C\mkern-8mu\vrule height1.45ex depth-.05ex
width.05em\mkern9mu\kern-.05em}
{\rm C\mkern-8mu\vrule height1ex depth-.07ex
width.035em\mkern9mu\kern-.035em}
{\rm C\mkern-8mu\vrule height.65ex depth-.1ex
width.025em\mkern8mu\kern-.025em}}}

\def\RR{{\rm I\kern-1.6pt {\rm R}}}

\def\ZZ{{\rm Z}\kern-3.8pt {\rm Z} \kern2pt}
\def\IB{\relax{\rm I\kern-.18em B}}
\def\ID{\relax{\rm I\kern-.18em D}}
\def\II{\relax{\rm I\kern-.18em I}}
\def\IP{\relax{\rm I\kern-.18em P}}

\newcommand{\bear}{\begin{eqnarray}}
\newcommand{\eear}{\end{eqnarray}}

\def\to{\rightarrow}

\def\to{\rightarrow}

\newcommand{\dd}{\mathrm{d}}

\def\d{\delta}

\def\f{\phi} 

\def\j{\psi}

 \def\th{\theta} 
\def\r{\rho} 

\def\6{\partial}

\def\atanh{{\rm arctanh}}

\def\ft#1#2{{\textstyle{{\scriptstyle #1}\over {\scriptstyle #2}}}}
\def\fft#1#2{{#1 \over #2}}
\def\del{\partial}
\def\sst#1{{\scriptscriptstyle #1}}

\def\dalemb#1#2{{\vbox{\hrule height .#2pt
 \hbox{\vrule width.#2pt height#1pt \kern#1pt
 \vrule width.#2pt}
 \hrule height.#2pt}}}

\def\0{{\sst{(0)}}}
\def\1{{\sst{(1)}}}
\def\2{{\sst{(2)}}}
\def\3{{\sst{(3)}}}
\def\4{{\sst{(4)}}}
\def\5{{\sst{(5)}}}
\def\6{{\sst{(6)}}}
\def\7{{\sst{(7)}}}
\def\8{{\sst{(8)}}}

   \let\d=\delta 
    
     \let\r=\rho
   \let\f=\phi

\def\bd{\begin{document}} \def\ed{\end{document}}
\let\Br=\Bigr \let\Bl=\Bigl 
\let\bm=\bibitem
\let\na=\nabla
\let\pa=\partial \let\ov=\overline 
\def\ba{\begin{eqnarray}}
\def\ea{\end{eqnarray}}
\def\ft#1#2{{\textstyle{{\scriptstyle #1}\over {\scriptstyle #2}}}}
\def\fft#1#2{{#1 \over #2}}
\def\del{\partial}
\def\sst#1{{\scriptscriptstyle #1}}
\def\oneone{\rlap 1\mkern4mu{\rm l}}
\def\ie{{\it i.e.\ }}
\def\via{{\it via}}
\def\semi{{\ltimes}}
\def\str{{\rm str}}
\def\jm{{\rm j}}
\def\im{{\rm i}}
\def\mapright#1{\smash{\mathop{-\!\!\!-\!\!\!-\!\!\!-\!\!\!-\!\!\!
 \longrightarrow}\limits^{#1}}}
\def\maprightt#1#2{\smash{\mathop{-\!\!\!-\!\!\!-\!\!\!-\!\!\!-\!\!\!
 \longrightarrow}\limits^{#1}_{#2}}}

\newcommand{\ho}[1]{$\, ^{#1}$}
\newcommand{\hoch}[1]{$\, ^{#1}$}
\newcommand{\ra}{\rightarrow}
\newcommand{\lra}{\longrightarrow}
\newcommand{\Lra}{\Leftrightarrow}
\newcommand{\bp}{\tilde \beta^\prime}
\newcommand{\Tr}{{\rm Tr} } 
\def\rme{{\rm e}}

\newfont{\namefont}{cmr10}
\newfont{\addfont}{cmti7 scaled 1440}
\newfont{\boldmathfont}{cmbx10}
\newfont{\headfontb}{cmbx10 scaled 1728}
\newcommand{\hyph}[1]{$#1$\nobreakdash-\hspace{0pt}}
\providecommand{\abs}[1]{\lvert#1\rvert}
\newcommand{\Nugual}[1]{$\mathcal{N}= #1 $}
\newcommand{\sub}[2]{#1_\text{#2}}
\newcommand{\partfrac}[2]{\frac{\partial #1}{\partial #2}}
\newcommand{\bsp}[1]{\begin{equation} \begin{split} #1 \end{split} \end{equation}}
\newcommand{\calF}{\mathcal{F}}
\newcommand{\calO}{\mathcal{O}}
\newcommand{\calM}{\mathcal{M}}
\newcommand{\calV}{\mathcal{V}}
\newcommand{\bbZ}{\mathbb{Z}}
\newcommand{\bbC}{\mathbb{C}}
\newcommand{\cK}{{\cal K}}
\newcommand{\Thq}{\Theta\left(\r-\r_q\right)}
\newcommand{\Dq}{\d\left(\r-\r_q\right)}
\newcommand{\kten}{\kappa^2_{\left(10\right)}}
\newcommand{\pbi}[1]{\imath^*\left(#1\right)}
\newcommand{\tth}{\tilde{\th}}
\newcommand{\tf}{\tilde{\f}}
\newcommand{\tj}{\tilde{\j}}
\newcommand{\tw}{\tilde{\omega}}
\newcommand{\tz}{\tilde{z}}
\newcommand{\prj}[2]{(\partial_r{#1})(\partial_{\j}{#2})-(\partial_r{#2})(\partial_{\j}{#1})}
\def\atanh{{\rm arctanh}}
\def\sech{{\rm sech}}
\def\csch{{\rm csch}}
\allowdisplaybreaks[1]
\def\red{\textcolor[rgb]{0.98,0.00,0.00}}
\newcommand{\Dan}[1] {{\textcolor{blue}{#1}}}
\numberwithin{equation}{section}
\renewcommand{\theequation}{{\rm\thesection.\arabic{equation}}}

\begin{document}
\baselineskip=15.5pt
\pagestyle{plain}
\setcounter{page}{1}
\begin{titlepage}
\begin{center}
\vskip .5in 
\noindent

{\Large \bf{Holographic Entanglement Entropy in Quiver Theories} } \\
\bigskip\medskip
Dimitrios Chatzis,$^\dagger$\footnote{\href{mailto:dchatzis@proton.me}{dchatzis@proton.me}} 
Ali Fatemiabhari,$^*$\footnote{\href{mailto:alifatemiabhari@gmail.com}{alifatemiabhari@gmail.com}} 
Mauro Giliberti$^\ddagger$\footnote{\href{mailto:mauro.giliberti@unifi.it}{mauro.giliberti@unifi.it}} and Madison Hammond$^\dagger$\footnote{\href{mailto:m.hammond.2412736@swansea.ac.uk}{m.hammond.2412736@swansea.ac.uk}}\\
\bigskip\medskip
{\small 
$^\dagger$ Department of Physics, Swansea University,\\Swansea SA2 8PP, United Kingdom\\
$^*$ Institute for Theoretical and Mathematical Physics, Lomonosov Moscow State University,\\119991 Moscow,
Russia\\
$^\ddagger$ Dipartimento di Fisica e Astronomia, Università degli Studi di Firenze,\\Via G. Sansone 1, I-50019 Sesto Fiorentino (Firenze), Italy
}

\vskip .5cm 
\vskip .9cm 
 	{\bf Abstract }\vskip .1in
\end{center}

\noindent
This work presents a study of the entanglement entropy (EE) in a class of four-dimensional ${\cal N}=1$ linear quiver SCFTs deformed by the presence of a VEV. We review the holographic backgrounds dual to these theories, and calculate the EE for different Ryu-Takayanagi embeddings. We allow the embeddings to explore, in addition to the usual spatial direction, the internal coordinate $z$, associated with the quiver degrees of freedom. Via the numerical optimization of splines on triangulations, we find the minimal configuration and the value of the EE for the different embeddings and quiver parameters. Our results agree with previous studies showcasing phase transitions in the EE. 
We also provide novel results illuminating the dependence of the EE on the fundamental and gauge degrees of freedom, signaling partial deconfinement, which are worthy of further study.
\vskip .5cm
\vskip .5cm
\vfill
\eject

\end{titlepage}

\setcounter{footnote}{0}

\small{
\tableofcontents}

\vspace{.5cm}

\hrule

\normalsize

\newpage
\renewcommand{\theequation}{{\rm\thesection.\arabic{equation}}}
%

\section{Introduction}

Building on the Maldacena conjecture and its extensions \cite{Maldacena:1997re, Gubser:1998bc, Witten:1998qj}, holography became a powerful tool for studying non-conformal field theories at strong coupling, with key contributions found in, e.g., \cite{Itzhaki:1998dd, Witten:1998zw, Boonstra:1998mp, Girardello:1999hj, Polchinski:2000uf}.
Following these advances, holographic methods were applied to confining field theories, with two primary strategies emerging. The first employs wrapped brane setups, as seen in \cite{Witten:1998zw, Maldacena:2000yy, Atiyah:2000zz, Edelstein:2001pu, Maldacena:2001pb}. The second centers on a certain two-node quiver theory with quasi-marginal deformations, realized in string theory through D3 and D5 brane dynamics on the conifold \cite{Klebanov:1998hh, Klebanov:2000nc, Klebanov:2000hb, Gubser:2004qj}. Connections between these two approaches have been established in works such as \cite{Maldacena:2009mw, Gaillard:2010qg, Caceres:2011zn, Elander:2011mh}.

The inclusion of dynamical quarks --- fields transforming in the fundamental representation of the gauge group --- poses a technically difficult problem. Despite this, advances have been achieved in various works, such as \cite{Casero:2006pt, Paredes:2006wb, Burrington:2007qd, Casero:2007jj, Bigazzi:2008gd, Hoyos-Badajoz:2008znk, Bigazzi:2008ie, Bigazzi:2008qq, Bigazzi:2009bk, Nunez:2010sf, Benini:2006hh, Benini:2007gx, Bigazzi:2014qsa, Bigazzi:2011it, Bea:2013jxa}. A defining feature of these setups is that the flavor branes (the gravitational duals of quark sources) are either extended or uniformly distributed (smeared) throughout the internal space. Consequently, instead of an original \(SU(N_f)\) flavor symmetry, the theory undergoes symmetry breaking \(SU(N_f) \to U(1)^{N_f}\) due to the VEVs of fields. This leads to notable dynamical effects, particularly the appearance of a singularity in the small-\(r\) (IR) region of the background when quarks are massless, making near-IR physics untrustworthy in such models --- there are certain exceptional cases without singularity mentioned in \cite{Filippas:2019puw}. Another aspect being the \textit{screening effect}, manifested as a probe string breaking, 
is formally suppressed by \( g_s \sim 1/N_c \) in the holographic regime. However, studies such as \cite{Bigazzi:2008gd, Bigazzi:2008ie, Bigazzi:2008qq, Bigazzi:2009gu} show that screening still leaves detectable imprints on observables despite this parametric suppression.

Furthermore, these models exhibit an undesirable characteristic: the inclusion of numerous flavor degrees of freedom typically results in a poorly-defined ultraviolet regime that lacks proper QFT definition. This pathological UV completion complicates the implementation of holographic renormalization procedures \cite{Papadimitriou:2004ap}.

In this work, we review and work with a holographic background that has resolved the aforementioned issues. The model builds upon the class of Anabalón-Ross type solutions \cite{Anabalon:2021tua, Anabalon:2022aig, Anabalon:2024che, Anabalon:2024qhf}, recently extended to holographic duals of confining QFTs and linear quivers with gapped infrared regimes in \cite{Nunez:2023nnl, Nunez:2023xgl, Fatemiabhari:2024aua, Chatzis:2024top, Chatzis:2024kdu, Barbosa:2024smw, Kumar:2024pcz}. These constructions provide infinite families of backgrounds that exhibit:
(i) proper field-theoretic UV behavior (manifested through asymptotic AdS$_5$ geometry),
(ii) well-regulated IR physics (represented by smooth geometries), and
(iii) fully backreacted localized sources while preserving four supercharges.

Within this framework, we compute entanglement entropy (EE) values for various gauge group choices in linear quiver QFTs. Our analysis reveals how quiver gauge nodes and localized sources modify the EE behavior, and we systematically investigate these effects across different quiver configurations.

Our analysis reveals a particularly noteworthy behavior of the Ryu-Takayanagi (RT) entropy: its embedding takes a nontrivial profile in the bulk geometry not solely along the conventional radial coordinate $r$ (encoding energy scales), but also along the $z$-direction, representing the linear quiver dimension. The Wilson loops in the aforementioned background had been studied in ref.~\cite{Giliberti:2024eii}. See also ref.~\cite{Das:2022njy} for other studies of RT surfaces wrapping internal directions.

In what follows, we outline the key concepts underlying this work, provide technical details of the system under investigation, and present the structure and objectives of our study.

We aim to study the effect of confinement and screening on the EE. It has been proposed in ref.~\cite{Klebanov:2007ws} the presence of a phase transition in the behavior of EE, due to the confining nature of the QFT --- see also critical analysis in \cite{Kol:2014nqa, Jokela:2020wgs}. Effects of flavor degrees of freedom have not been studied before, and a cross-over between confinement and screening could affect the dependence of EE on the order parameter of the mentioned phase transition. More interestingly, there is the possibility of partial deconfinement \cite{Hanada:2016pwv} that may apply to the quivers studied here. Some nodes of the quiver degrees of freedom can transition between confinement/deconfinement, while others are stable. This case will be considered in our analysis. 

We employ holography as our primary investigative tool. While a precise holographic dual to QCD remains elusive (as discussed earlier), we focus on a well-defined holographic description of a family of four-dimensional, balanced linear quivers. These theories preserve four supercharges, exhibit a strongly coupled conformal UV fixed point, and inherently incorporate fundamental matter. The dual gravitational setup features gauge groups with adjoint and bifundamental matter, as well as matter in the fundamental representation of certain gauge groups. Holographically, the fundamental matter is introduced through localized D-brane sources present in the bulk geometry.

We calculate the EE using the RT method \cite{Ryu:2006bv}. We divide the space into two subregions in the boundary of the background where the dual UV QFT lives. We find the minimal area surface that has its boundary ending on the boundary of the two mentioned subregions. The surface evolves inside the bulk, `falling' towards the end of radial $r$-coordinate, with a nontrivial profile in the `quiver tail' direction (chosen to be the $z$ direction in the body of the paper).

The minimization of the probe surface's area poses analytical challenges, making it impossible to find a closed-form solution. We therefore adopt a numerical approach, the details of which are presented in subsequent sections.

We plot the value of the entropy of the probe $S_\text{EE}$ with respect to the partition lengths in the QFT side  $L$ and $z_\star$ (in the $x$ and $z$ directions, respectively); and we provide surface profiles, confirming the dependence on the radial coordinate and the quiver direction ($r$ and $z$, respectively). We also observe some evidence for partial deconfinement, as is described in the following sections.

The paper is organized as follows.
In section~\ref{section-quiver-QFT}, we present a family of holographic backgrounds used in our analysis. This section provides the geometric basis for our subsequent investigations.
Section~\ref{section-EE} develops the theoretical framework for analyzing EE in our setup. We derive the complete EE integral functional for a volume having a nontrivial profile both in the radial holographic direction $r$ and the quiver direction $z$.
The numerical implementation and results are presented in section~\ref{sec-numerical}, where we detail our computational approach and analyze the obtained solutions. 
We conclude in section~\ref{concl} with a summary of our results and a discussion of promising directions for future research. 

\section{Background and dual QFT}\label{section-quiver-QFT}

We begin by considering the class of supergravity backgrounds constructed in \cite{Apruzzi:2013yva,afprt2015,ct2015}, and further elaborated in \cite{Chatzis:2024top,Chatzis:2024kdu,Giliberti:2024eii,Fatemiabhari:2024aua}. These are an infinite family of backgrounds dual to non-Lagrangian SCFTs, and can be interpreted as twisted compactifications of a D6-D8-NS5 system on a negative curvature Riemann surface $\Sigma$.

The procedure for obtaining the background closely follows ref.~\cite{Giliberti:2024eii}: first, we identify the holographic duals of six-dimensional $\mathcal{N}=(1,0)$ linear quiver SCFTs, commonly denoted as $\mathrm{SCFT}_6$. These theories have a bosonic global symmetry group $SO(2,6) \times SU(2)_R$, where $SO(2,6)$ is the six-dimensional conformal group associated with the eight Poincaré supercharges. Then we compactify the $\mathrm{SCFT}_6$ on a hyperbolic manifold, flowing at low energies to a family of four-dimensional $\mathcal{N}=1$ SCFTs to give a family of $\mathrm{AdS}_5$ solutions dual to $4D\; \mathcal{N}=1$ SCFTs \cite{bpt2017,Apruzzi:2015zna}. The flows were constructed in \cite{Merrikin:2022yho}.

The corresponding massive type IIA supergravity backgrounds are described by a metric, a dilaton $\Phi$, an NS two-form $B_2$, and Ramond-Ramond fields $F_0$ and $F_2$, where $F_0$ is a piecewise constant function determined by the structure of $\alpha(z)$. The geometry is encoded in a continuous, piecewise-cubic function $\alpha(z)$, which determines supersymmetric solutions to the equations of motion of massive type IIA supergravity with mass parameter $F_0$.

The equations of motion for massive IIA are satisfied when
\begin{equation}
    \dddot{\alpha}=-162\pi^3 F_0;
\end{equation}
this was checked in \cite{Merrikin:2022yho}. 
Since $F_0$ is a piecewise constant function --- discontinuous at the locations of D8-brane sources but otherwise constant --- the solution for $\alpha(z)$ is a continuous, piecewise-cubic function:
\begin{equation}
    \alpha(z)=a_0 +a_1 z + \frac{a_2}{2}z^2 -\frac{162\pi^3 F_0}{6} z^3.
\end{equation}
This function determines the background geometry and provides a holographic dual to the six-dimensional $\mathcal{N} = (1,0)$ quiver SCFT at the origin of its tensor branch.

The information of the dual quiver field theory, including the ranks of the gauge groups, is encoded in the so-called rank function $\mathcal{R}(z)$, which is related to $\alpha(z)$ via
\begin{equation}
\mathcal{R}(z)=-\frac{1}{81\pi^2}\ddot{\alpha}(z)= \left\{\begin{array}{lr}
        N_1 z & 0\leq z \leq 1 \\
        N_1 + (N_2 -N_1)(z-1) & 1\leq z \leq 2 \\
        N_k + (N_{k+1}-N_k)(z-k) & k\leq z \leq (k+1) \\
        \dots \\
        N_{P-1}(P-z) & (P-1)\leq z \leq P. 
\end{array}\right.
\end{equation}
This function is continuous and piecewise-linear, with its second derivative given by a sum of delta functions reflecting the $SU(F_k)$ gauge group ranks
\begin{equation}
\mathcal{R}^{\prime\prime}(z)= \sum_{k=1}^{P-1} F_k \delta(z-k).
\end{equation}

We can find $\alpha(z)$ by integrating twice and imposing the boundary conditions $\alpha(0)=\alpha(P)=0,$
\begin{equation}
\ddot{\alpha}=-81\pi^2 \mathcal{R}(z).
\end{equation}

The 6D SCFTs arise as UV completions of linear quiver gauge theories. 

\begin{figure}[htp]
\begin{center}
	\begin{tikzpicture}
	\node (1) at (-4,0) [circle,draw,thick,minimum size=1.4cm] {N$_1$};
	\node (2) at (-2,0) [circle,draw,thick,minimum size=1.4cm] {N$_2$};
	\node (3) at (0,0)  {$\dots$};
	\node (5) at (4,0) [circle,draw,thick,minimum size=1.4cm] {N$_{P-1}$};
	\node (4) at (2,0) [circle,draw,thick,minimum size=1.4cm] {N$_{P-2}$};
	\draw[thick] (1) -- (2) -- (3) -- (4) -- (5);
	\node (1b) at (-4,-2) [rectangle,draw,thick,minimum size=1.2cm] {F$_1$};
	\node (2b) at (-2,-2) [rectangle,draw,thick,minimum size=1.2cm] {F$_2$};
	\node (3b) at (0,0)  {$\dots$};
	\node (5b) at (4,-2) [rectangle,draw,thick,minimum size=1.2cm] {F$_{P-1}$};
	\node (4b) at (2,-2) [rectangle,draw,thick,minimum size=1.2cm] {F$_{P-2}$};
	\draw[thick] (1) -- (1b);
	\draw[thick] (2) -- (2b);
	\draw[thick] (4) -- (4b);
	\draw[thick] (5) -- (5b);
	\end{tikzpicture}
\end{center}
\caption{\small Quiver diagram for a linear quiver with $(P-1)$ gauge nodes. Balancing the quiver means that we have $F_k= 2N_k - N_{k-1} - N_{k+1}$ for each node.}\label{quiver_fig}
\end{figure}
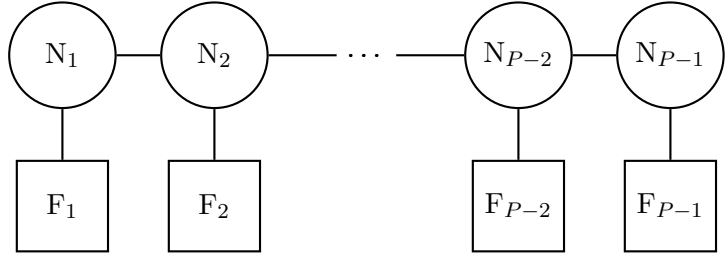

The final step in our construction, as described in \cite{Giliberti:2024eii}, involves breaking conformal invariance and flowing to a gapped theory. This is achieved by implementing a compactification on a two dimensional Riemann surface followed by a twisted compactification on a circle, in a manner that preserves four supercharges, following the procedure outlined in \cite{Chatzis:2024top,Chatzis:2024kdu}. The metric and dilaton are given by
\begin{align}\label{AFPRT_metric}
   \mathrm{d}s_{10}^2=&18\pi\sqrt{-\frac{\alpha}{6 \ddot{\alpha}}}\left[\mathrm{d}s_5^2+\frac{1}{3} \mathrm{d}s_\Sigma^2-\frac{\ddot{\alpha}}{6\alpha}\mathrm{d}z^2-\frac{\alpha \ddot{\alpha}}{6 \dot{\alpha}^2-9 \alpha \ddot{\alpha}} \left(\mathrm{d}\theta^2+\sin^2 \theta \mathcal{D}\psi^2\right)
    \right], \\
    &\mathrm{d}s_\Sigma^2= \frac{4(\mathrm{d}v_1^2+\mathrm{d}v_2^2)}{(1- v_1^2-v_2^2)^2},\nonumber
    \\
    & {\cal D}\psi= \mathrm{d}\psi - 3\mathcal{A} - A_\Sigma, \qquad 
     \mathcal{A}=q\left(\frac{1}{r^2}-\frac{1}{r_*^2} \right)  \mathrm{d}\phi, \qquad A_\Sigma= \frac{2(v_1\mathrm{d}v_2- v_2 \mathrm{d}v_1)}{1-v_1^2-v_2^2}, \\
     & \mathrm{d}s_5^{2}=\frac{r^{2}}{l^{2}}(-\mathrm{d}t^{2}+\mathrm{d}x_1^{2}+\mathrm{d}x_2^{2}+f(r)\mathrm{d}\phi^2)+\frac{l^2\mathrm{d}r^{2}}{r^2f(r)}, \\
     & e^{-4\Phi}= \frac{1}{2^5 3^{17}\pi^{10}}\left( -\frac{\ddot{\alpha}}{\alpha}\right)^3 \left( 2\dot{\alpha}^2-3 \alpha \ddot{\alpha}\right)^2,\label{AFPRT_dilaton}
\end{align}
with $\alpha=\alpha(z)$, $f(r)=1-\frac{\mu}{r^4}-\frac{q^2l^2}{r^6}$. The parameter $r_*$ represents the smallest root of $f(r)$ and corresponds to the minimal value of the radial coordinate, at which the spacetime ends smoothly in a cigar-like fashion. 


A key conceptual point must also be addressed. While the low-energy regime of the six-dimensional mother SCFT (dual to the original gravity solution mentioned in ref.~\cite{Giliberti:2024eii}) is described by a Lagrangian, the compactification process yields a four-dimensional theory that is non-Lagrangian. Consequently, our terminology of placing gauge nodes at integer $z$ positions and flavor nodes at the kinks of the rank function is a useful heuristic rather than a precise description, as the gauge/flavor node paradigm may not hold strictly true in the resulting 4D QFT.

We will calculate the entanglement entropy in a QFT defined by a linear quiver gauge theory. We consider a family of linear quivers, i.e. a chain of gauge nodes labeled by the \textit{quiver direction}, a coordinate $z$ ranging from $0$ to $P$, with rank functions that vary as a function of $z$. The quiver has flavor (global symmetry) groups attached at certain nodes of integer $z$; these flavor groups are implemented in the dual geometry by localized sources (D-brane stacks) at those positions in $z$. The parameter $P$ sets the total length of the quiver. There are $P$ gauge nodes in total, and flavor groups can be attached at specific integer values of $z$. Unless otherwise specified, in what follows, we set $P=10$ to study the example of a quiver with nine gauge nodes and one flavor group.

We consider the following three quivers as examples, as in ref.~\cite{Giliberti:2024eii}.

\noindent\textbf{Quiver I: Scalene triangle}

The scalene triangle quiver has one flavor group, placed at the node $z=P-1$ as is shown in the quiver diagram:

\begin{center}
	\begin{tikzpicture}
	\node (1) at (-6,0) [circle,draw,thick,minimum size=1.2cm] {N};
	\node (2) at (-4,0) [circle,draw,thick,minimum size=1.2cm] {2N};
	\node (3) at (-2,0) [circle,draw,thick,minimum size=1.2cm] {3N};	
	\node (4) at (0,0) {$\dots$};
	\node (6) at (4,0) [rectangle,draw,thick,minimum size=1.2cm] {PN};
	\node (5) at (2,0) {(P-1)N};
	\draw[thick] (1) -- (2) -- (3) -- (4) -- (5)-- (6);
	\draw[thick] (2,0) circle (0.7cm) ;
	\draw[thick] (1,0) -- (1.3,0);
	\draw[thick] (2.7,0) -- (3.3,0);
	\end{tikzpicture}\
\end{center}

The function $\alpha(z)$ used is
\begin{equation}
    \alpha(z) = -\frac{81\pi^2}{6}\left\{
    \begin{array}{lr}
    (1-P^2)z + z^3 & 0\leq z \leq P-1\; \\
    (2P^2 -3P +1)(z-P) + (P-1)(P-z)^3 \qquad & P-1 \leq z \leq P,
    \end{array}\right.
\end{equation}
with rank function
\begin{equation}
    \mathcal{R}(z) = -\frac{1}{81\pi^2}\ddot{\alpha}(z) = \left\{
    \begin{array}{lr}
    z & 0 \leq z \leq P-1\;\\
    (P-1)(P-z) \qquad & P-1 \leq z \leq P.
    \end{array}\right.
\end{equation}

\begin{figure}[htp]
    \centering
    \includegraphics[width=0.8\linewidth]{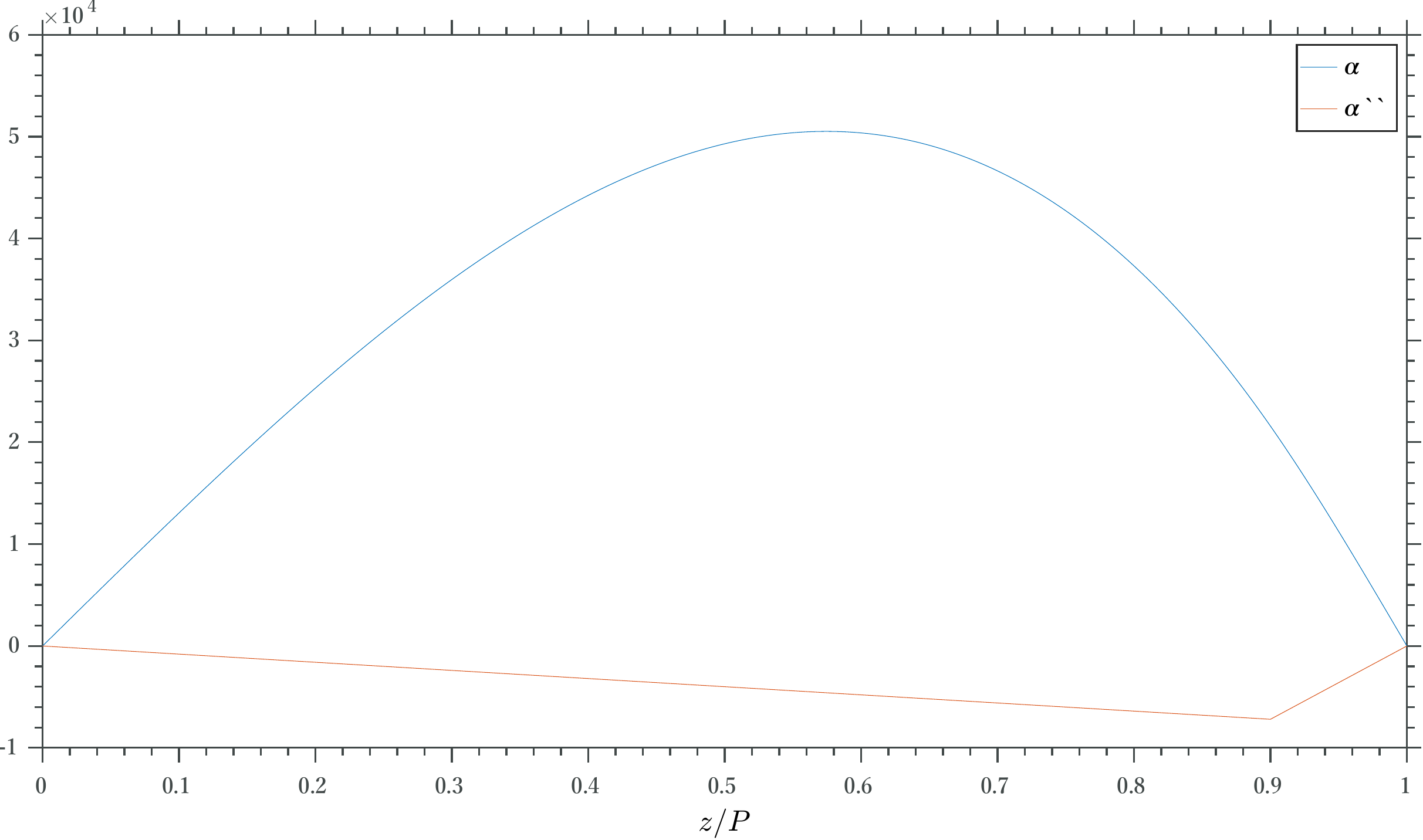}
    \caption{\small The rank-function increases linearly from $z=0$ to $z=P-1$, then goes to zero at $z=P$. The rank function profile is a non-symmetric triangle shape, since the slopes on either side of the kink differ.}
\end{figure}

\noindent\textbf{Quiver II: Isosceles triangle}

The isosceles triangle quiver has one flavor group placed at the central node, $z=\frac{P}{2}$ and the corresponding diagram is:

\begin{center}
	\begin{tikzpicture}
	\node (1) at (-6,0) [circle,draw,thick,minimum size=1.2cm] {N};
	\node (2) at (-4,0) [circle,draw,thick,minimum size=1.2cm] {2N};	
	\node (3) at (-2,0) {$\dots$};
 \node (4) at (0,0) [circle,draw,thick,minimum size=1.2cm] {PN/2};
 \node (5) at (2,0) {$\dots$};
	\node (6) at (4,0) {(P-1)N};
 \node (7) at (6,0) [circle,draw,thick,minimum size=1.2cm] {PN};
 \node (4q) at (0,-2) [rectangle,draw,thick,minimum size=1.2cm] {2N};
	\draw[thick] (1) -- (2) -- (3) -- (4) -- (5)-- (6) -- (7);
	\draw[thick] (4,0) circle (0.7cm) ;
	\draw[thick] (1,0) -- (1.3,0);
	\draw[thick] (2.7,0) -- (3.3,0);
 \draw[thick] (4) -- (4q);
	\end{tikzpicture}\
\end{center}

The function $\alpha(z)$ is given by
\begin{equation}
    \alpha(z) = -81\pi^2\left\{
    \begin{array}{lr}
   - \frac{P^2}{8}z + \frac{z^3}{6}& 0\leq z \leq \frac{P}{2} \;\\
    -\frac{P^2}{8}(P-z) + \frac{1}{6}(P-z)^3  \qquad & \frac{P}{2} \leq z \leq P,
    \end{array}\right.
\end{equation}
with rank function
\begin{equation}
    \mathcal{R}(z) = -\frac{1}{81\pi^2}\ddot{\alpha}(z) = \left\{
    \begin{array}{lr}
    z & 0 \leq z \leq \frac{P}{2}\;\\
    P-z \qquad & \frac{P}{2} \leq z \leq P.
    \end{array}\right.
\end{equation}

\begin{figure}[htp]
    \centering
    \includegraphics[width=0.8\linewidth]{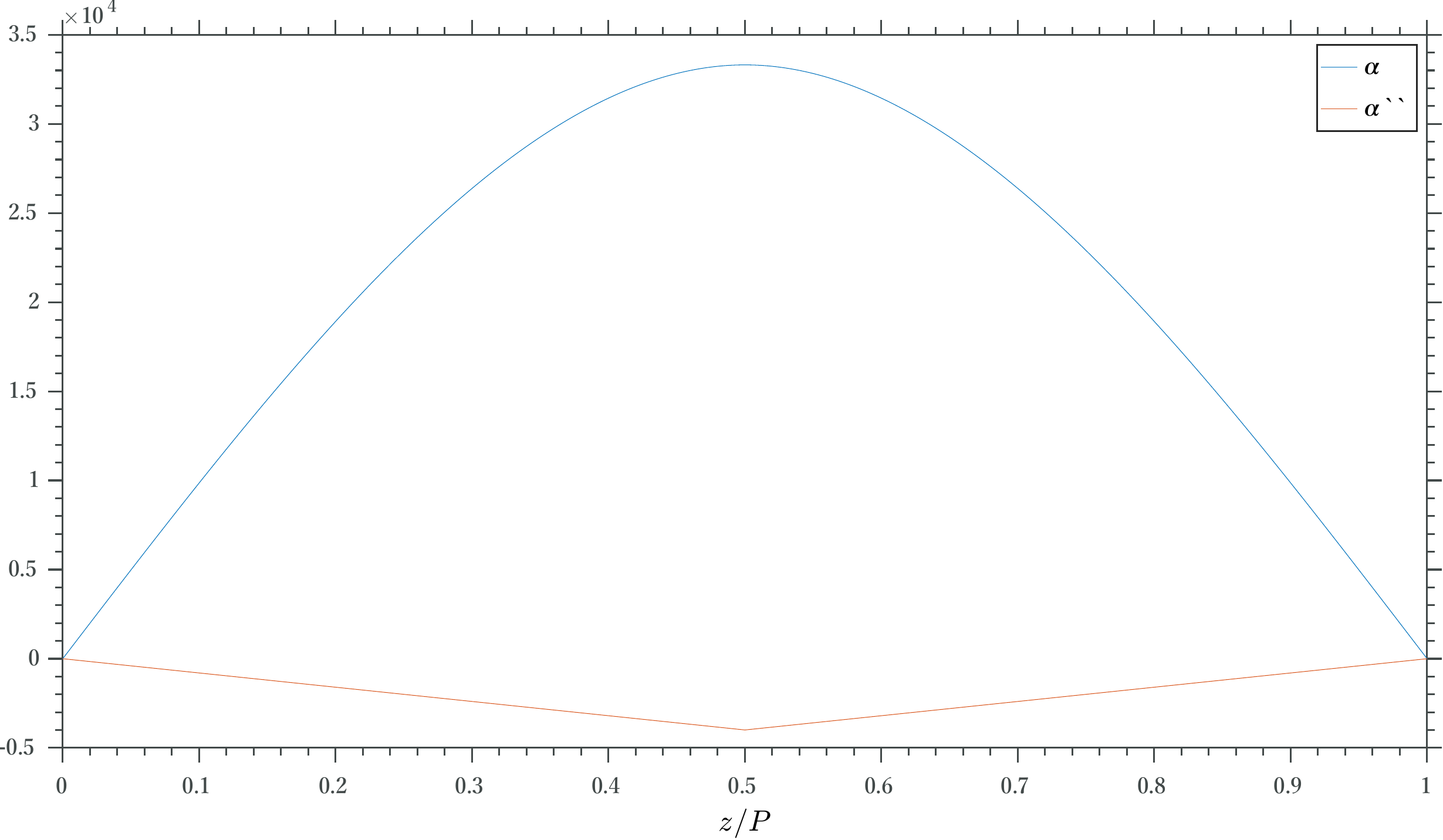}
    \caption{\small The rank function increases in magnitude linearly up to $z=\frac{P}{2}$, then decreases symmetrically after that. The rank function has equal magnitude slopes such that its plot is an isosceles triangle shape.}
\end{figure}

\noindent\textbf{Quiver III: Isosceles trapezoid}

The isosceles trapezoid quiver example has two flavor groups, one at each end of the quiver, $z=1$ and $z=P-1$:

\begin{center}
	\begin{tikzpicture}
 \node (0) at (-8,0) [rectangle,draw,thick,minimum size=1.2cm] {N};
	\node (1) at (-6,0) [circle,draw,thick,minimum size=1.2cm] {N};
	\node (2) at (-4,0) [circle,draw,thick,minimum size=1.2cm] {N};
	\node (3) at (-2,0) [circle,draw,thick,minimum size=1.2cm] {N};	
	\node (4) at (0,0) {$\dots$};
	\node (6) at (4,0) [rectangle,draw,thick,minimum size=1.2cm] {N};
	\node (5) at (2,0) [circle,draw,thick,minimum size=1.2cm] {N};
	\draw[thick] (0) -- (1) -- (2) -- (3) -- (4) -- (5)-- (6);
	\draw[thick] (1,0) -- (1.3,0);
	\draw[thick] (2.7,0) -- (3.3,0);
	\end{tikzpicture}\
\end{center}

In this example, we take
\begin{equation}
    \alpha(z) = -81\pi^2\left\{
    \begin{array}{lr}
    \frac{1}{2}(1-P)z + \frac{z^3}{6} & 0\leq z \leq 1 \\
    \frac{1}{6} - \frac{P}{2}z + \frac{z^2}{2} & 1 \leq z \leq P-1\;\\
    \frac{1}{2}(1-P)(P-z) + \frac{1}{6}(P-z)^3 \qquad & P-1 \leq z \leq P,
    \end{array}\right.
\end{equation}
with rank function
\begin{equation}
    \mathcal{R}(z) = -\frac{1}{81\pi^2}\ddot{\alpha}(z) = \left\{
    \begin{array}{lr}
    z & 0 \leq z \leq 1\;\\
    1 & 1\leq z \leq P-1\;\\
    P-z \qquad & P-1 \leq z \leq P.
    \end{array}\right.
\end{equation}

\begin{figure}[htp]
    \centering
    \includegraphics[width=0.8\linewidth]{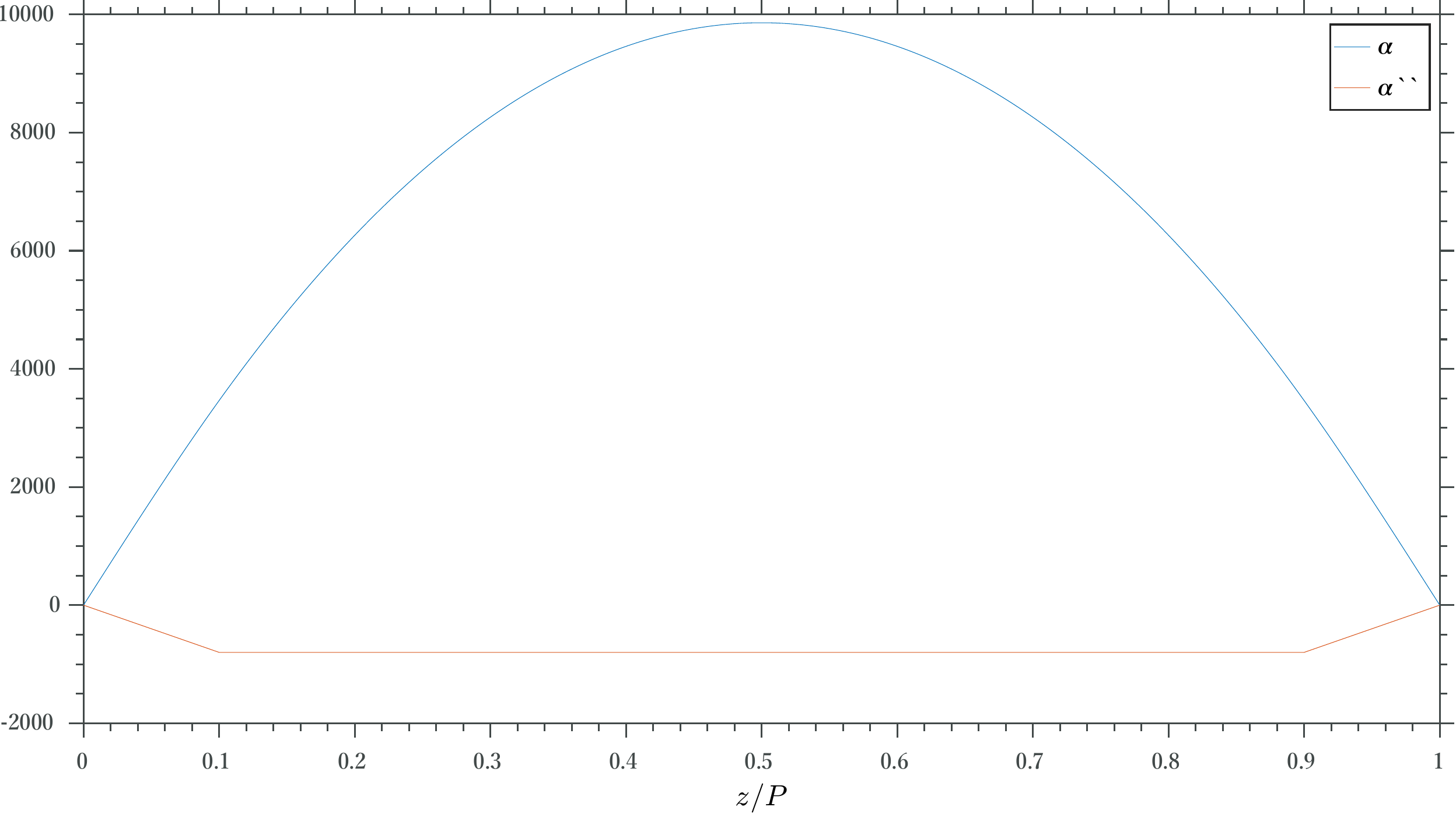}
    \caption{\small The rank function increases linearly from $z=0$ to $z=1$, is constant between the two flavor nodes, and then decreases linearly from $z=P-1$ to $z=P$.}
\end{figure}

\FloatBarrier
\section{Entanglement entropy}\label{section-EE}

Let us present the concept of holographic entanglement entropy, derived in the seminal work of \cite{Ryu:2006bv}. One of the many appealing aspects of this observable is its use as an order parameter for phase transitions describing confinement/deconfinement, with notable exceptions as alluded to earlier in the text. The detailed argument can be found in refs.~\cite{Klebanov:2007ws,Nishioka:2009un,Kol:2014nqa,Jokela:2020wgs}.

Consider a $\mathrm{CFT}_{d+1}$ dual to a gravity background that is asymptotically $\mathrm{AdS}_{d+2}$ and let $\cal D$ denote a subsystem of the $\mathrm{CFT}_{d+1}$ with $d$-dimensional boundary $\partial \cal D$. 
Then the entanglement entropy of $\cal D$ with respect to its complement is given by the von Neumann entropy $S_{\cal D}=-\mathrm{tr}_{\cal D}\log \rho_{\cal D}$, where $\rho_{\cal D}$ is the reduced density matrix for the subsystem, and it expresses the amount of entanglement between the two subsystems. In holography, the formula to compute the entanglement entropy is given in terms of a codimension-2 submanifold $\gamma_{\cal D}$ that is a {\it minimal surface} in the gravity and background whose boundary is $\partial\cal D$:
\begin{equation}\label{S_D}
    S_{\text{EE}}[\gamma_{\cal D}]=\frac{\text{Area}(\gamma_{\cal D})}{4 G_N^{(d+2)}},
\end{equation}
where $G_N^{(d+2)}$ is Newton's constant in $(d+2)$ dimensions. In this work, we will consider $\cal D$ to be a strip $[-L/2,L/2]\times\mathbb{R}$ that extends in a $\mathbb{R}^{1,1}$ subspace of the Minkowski space.

To compute the entanglement entropy of the infinite strip in the solution \eqref{AFPRT_metric}--\eqref{AFPRT_dilaton} we take the codimension two manifold to be spanned by the following coordinates: \begin{equation}\gamma_8[x_1,x_2,\phi,v_1,v_2,z,\theta,\psi],\end{equation}
with the strip extending in $[t,x_1]$. Then \eqref{S_D} reads (in string frame):

\begin{equation}\label{S_general}
\begin{split}
    S_{\text{EE}}&=\frac{1}{4G_N}\int_{\gamma_8}\mathrm{d}^8\sigma\sqrt{e^{-4\Phi}\mathrm{det}(g_{\gamma_8})},
\end{split}
\end{equation}
where $g_{\gamma_8}$ denotes the pullback of the metric on $\gamma_8$ and $G_N$ the ten-dimensional Newton constant.

In the following, we will present three ansätze for the radial coordinate: a function of the spatial coordinate $x_1$, a function of the quiver coordinate $z$, as well as a function of both $x_1$ and $z$. We have checked that all three embeddings are consistent, that is, they are solutions to the truncated equations of motion for $r=r(x_1,x_2,\phi,v_1,v_2,z,\theta,\psi)$. By allowing a nontrivial $z$ dependence on $r$, the probe explores the part of the internal space associated with the quiver tail, thus making accessible the study of entanglement between degrees of freedom living in different gauge groups in the dual $\mathrm{QFT}$.

\subsection{Case I: \texorpdfstring{$r(x_1)$}{r(x₁)}}\label{sec:rofx}

The simplest embedding one usually considers is the case where $r=r(x_1)$, which was covered in \cite{Chatzis:2024kdu}. There, it was shown that the entanglement entropy is double valued with respect to the spatial length $L$ along $x_1$, that is, it undergoes a phase transition: considering an embedding $\gamma_8$ which explores the bulk in a connected way, there is a value for $L$ after which a second embedding $\gamma_8^{\prime}$ is physically favorable consisting of two disconnected ``sheets" with boundaries at $x_1=\pm L/2$, expressing the confined phase with zero entanglement entropy. The induced metric and determinant which enters \eqref{S_general} are, in this case:

\begin{equation}
\begin{split}
    \mathrm{d}s_{\gamma_8}^2= 18\pi\sqrt{-\frac{\alpha}{6\ddot{\alpha}}}&\biggl\{  \frac{r^2}{l^2}\left[ \left(1+ \frac{l^4 r'^2}{r^4 f(r)} \right)\mathrm{d}x_1^2+ \mathrm{d}x_2^2+ f(r)\mathrm{d}\phi^2 \right] 
+\frac{1}{3} \mathrm{d}s_\Sigma^2-\frac{\ddot{\alpha}}{6\alpha}\mathrm{d}z^2\\
 & -\frac{\alpha \ddot{\alpha}}{6 \dot{\alpha}^2-9 \alpha \ddot{\alpha}} \left(\mathrm{d}\theta^2+\sin^2 \theta \mathcal{D}\psi^2\right)
\biggr\},
\end{split}
\end{equation}

\begin{equation}
    \mathrm{det(g_{\gamma_8})}=-\frac{279936\pi^8\alpha^5\sin^2\theta}{l^6(v_1^2+v_2^2-1)^4\ddot{\alpha}(2\dot{\alpha}^2-3\alpha\ddot{\alpha})^2}\Big[ r^6f(r) + l^4 r^2 (\partial_{x_1}r)^2\Big];
\end{equation}
and the action\footnote{Let us note that the \textit{action} of the minimal Ryu-Takayanagi surface and the entanglement entropy \textit{area functional} are in fact the same, so we will use these words interchangeably in the rest of the paper.} reads 
\begin{equation}\label{SEE_rx1}
\begin{split}
&S_{\mathrm{EE}}={\cal N}_{x_1}\int _{-L/2}^{L/2}\mathrm{d}x_1\sqrt{F_{x_1}^2(r)+G_{x_1}^2(r)(\partial_{x_1}r)^2},\\
&{\cal N}_{x_1}=\frac{L_{x_2}L_{\phi}\mathrm{Vol}(\Sigma)}{486 \pi G_N}\int_0^P \mathrm{d}z \left(-\alpha\ddot{\alpha}\right),\quad F^2_{x_1}(r)=\frac{r^6f(r)}{l^6},\quad G_{x_1}^2(r) =\frac{r^2}{l^2},
\end{split}
\end{equation}
where $L_{x_2}=\int\mathrm{d}x_2$, $L_{\phi}$ is the period of $\phi$ and ${\cal N}_{x_1}$ is a numerical factor associated with the central charge of the $\mathrm{UV}$ $\mathrm{CFT}$. We can use the Euler-Lagrange equations for the Lagrangian of \eqref{SEE_rx1}, which yields the following conserved Hamiltonian: 
\begin{equation} \label{eq:Hr0}
    H=\frac{\delta\mathcal{L}_{\mathrm{EE}}}{\delta (\partial_{x_1}r)}\partial_{x_1}r - \mathcal{L}_{\mathrm{EE}}=-\frac{F_{x_1}^2(r)}{\sqrt{F_{x_1}^2(r)+G_{x_1}^2(r)(\partial_{x_1}r)^2}}=F_{x_1}(r_0).
\end{equation}
We set it to the constant value $F_{x_1}(r_0)$, with $r_0$ denoting the turning point of the embedding. The above implies
\begin{equation}
    \frac{\mathrm{d}r}{\mathrm{d}x_1}=\pm \frac{F_{x_1}(r)}{G_{x_1}(r)F_{x_1}(r_0)}\sqrt{F_{x_1}^2(r)-F_{x_1}^2(r_0)},
\end{equation}
from which we can obtain an integral expression for the length of the strip, in terms of the turning point \cite{Kol:2014nqa}: 
\begin{equation}
    L_{\mathrm{EE}}(r_0)=\int_{-L_{\mathrm{EE}}/2}^{L_{\mathrm{EE}}/2}\mathrm{d}x_1=2F_{x_1}(r_0)\int_{r_0}^{\infty}\mathrm{d}r\frac{G_{x_1}(r)F_{x_1}(r)}{\sqrt{F_{x_1}^2(r)-F_{x_1}^2(r_0)}}.\label{Lofr0_case1}
\end{equation}

We also use a regulator for the entanglement entropy action, which amounts to subtracting the disconnected configuration of the two embeddings extending from $x_1=\pm L/2$ at $r=\infty$ down to $r=r_*$: 
\begin{eqnarray}
    S_{\mathrm{EE}}= \mathcal{N}_{x_{1}} \left[\int_{-L/2}^{L/2}\mathrm{d}x_1\sqrt{F^2_{x_1}+G^2_{x_1}(\partial_{x_1}r)^2}- 2\int_{r_*}^{\infty}\mathrm{d}r\,G_{x_1}\right].\label{regularized_rx_action}
\end{eqnarray}

\begin{figure}[htp]
    \begin{center}
    	\includegraphics[width=0.44\textwidth]{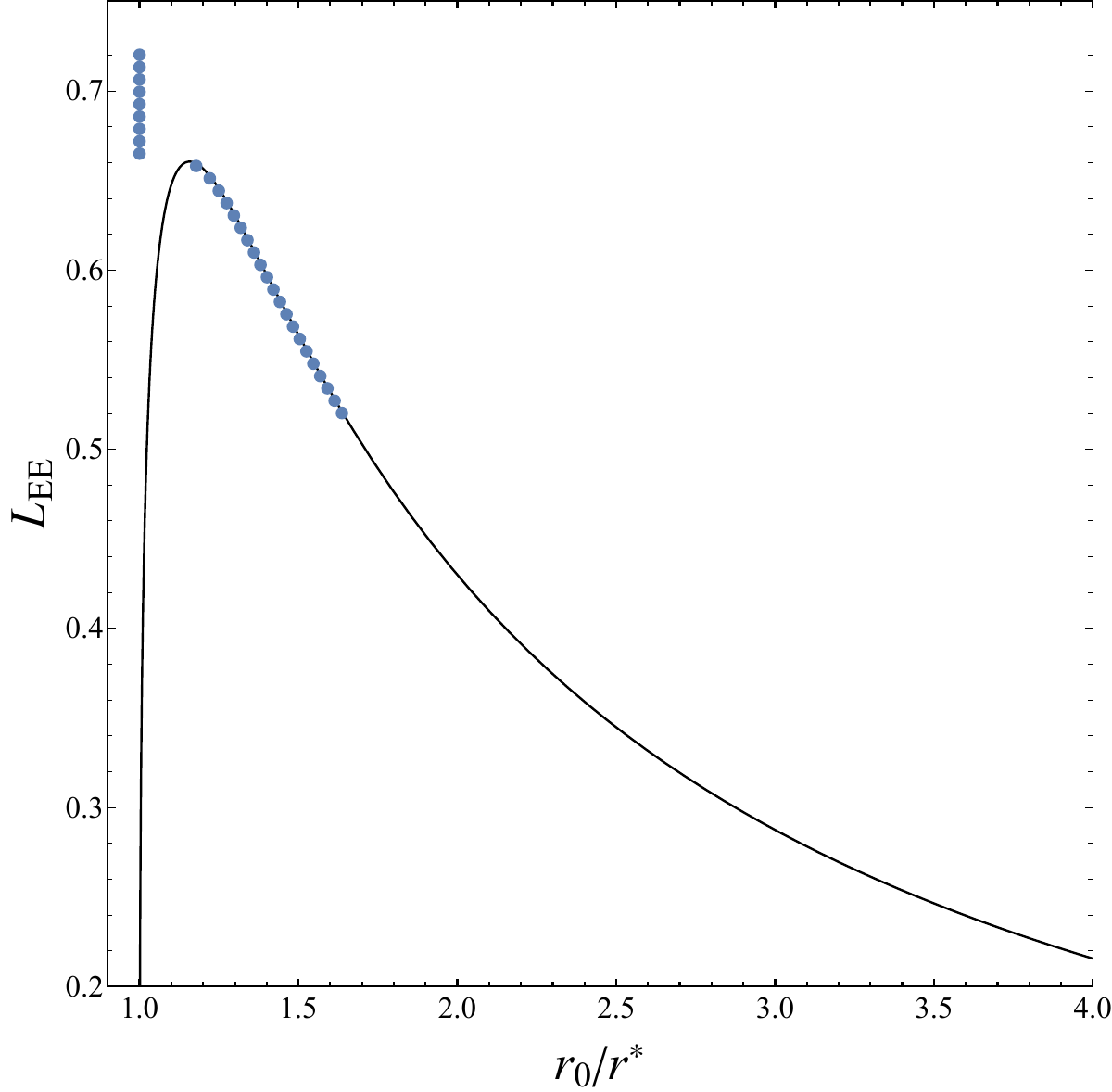}
        \includegraphics[width=0.45\textwidth]{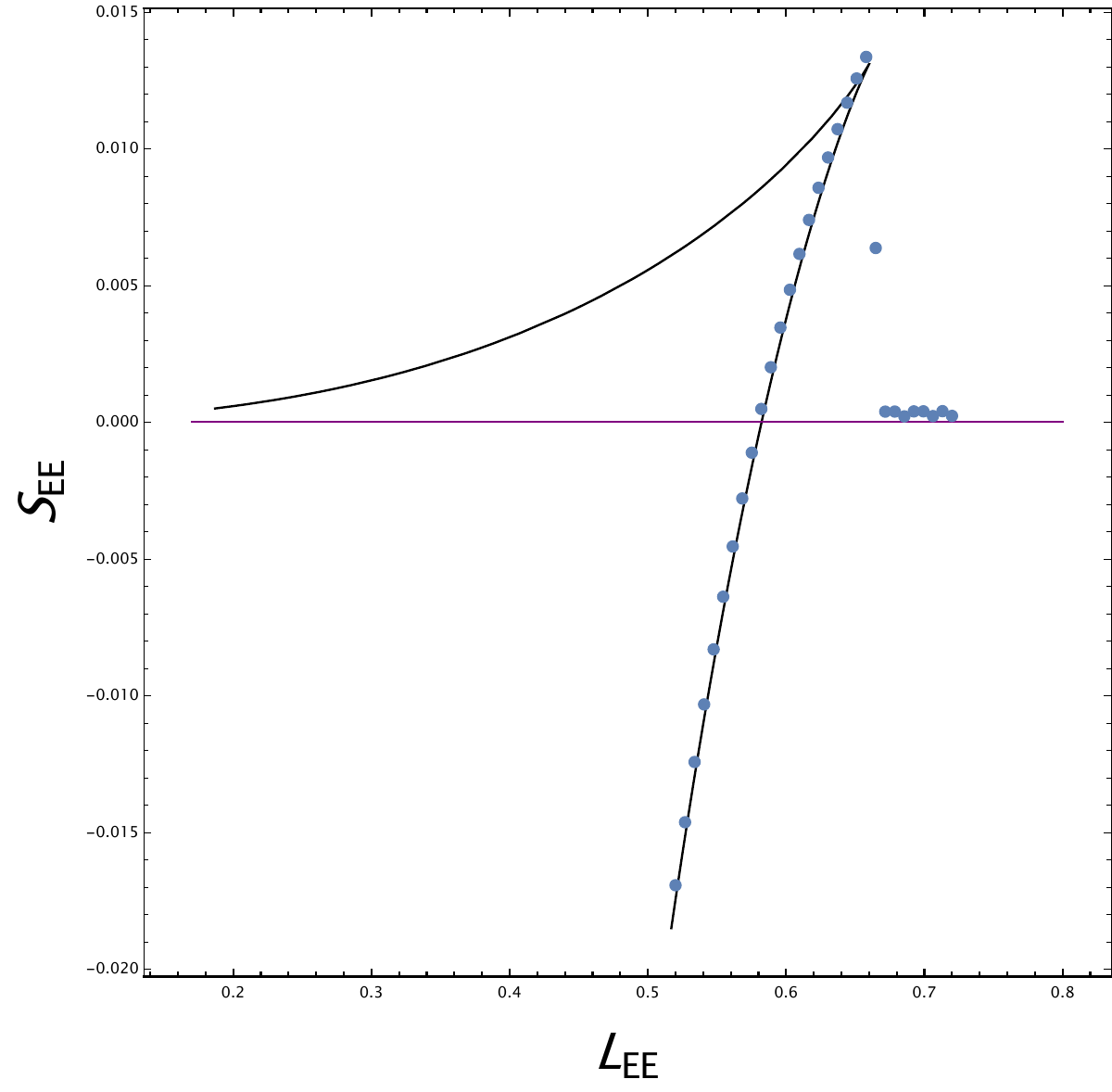}
    \end{center}
    \caption{\small Comparison of results obtained from numerical integration and numerical optimization for case I, presented in \ref{sec:rofx}. In black, the numerical integration result for the strip length \eqref{Lofr0_case1} as a function of the $r$-value of the turning point (left) and the entanglement entropy \eqref{SEE_rx1} as a function of the strip length (right). In blue, the points resulting from the numerical optimization of $S_\text{EE}(r)$, for different values of $L_\text{EE}$; it can be seen how for $L_\text{EE}>L_\text{EE}^\text{crit,1}\simeq0.66$ the points lie on a vertical line $r_0=r_*$, that we regard as the disconnected configuration.}
\label{fig:rx_case1}
\end{figure}

\begin{figure}[htp]
    \begin{center}
    	\includegraphics[width=0.44\textwidth]{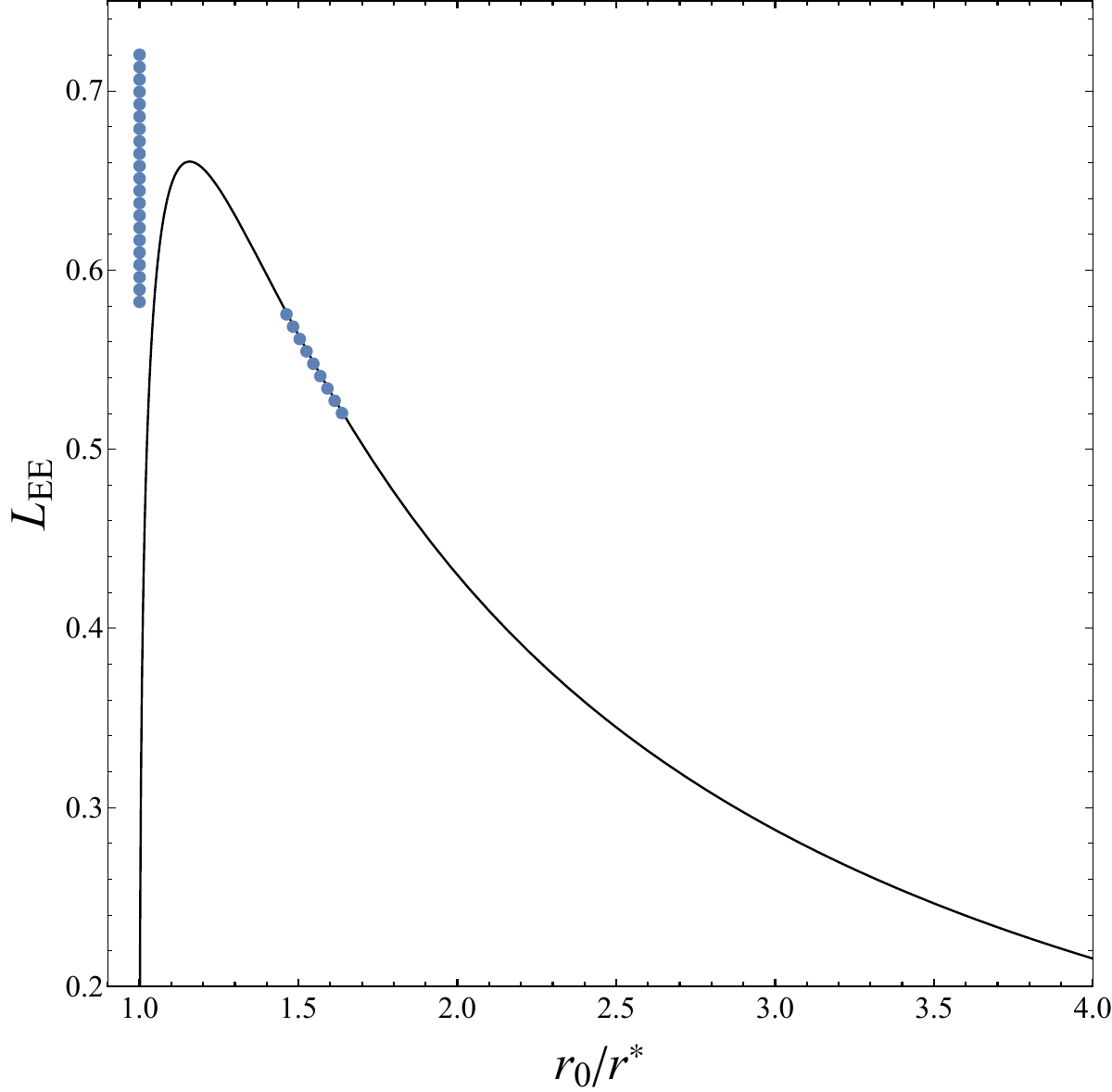}
        \includegraphics[width=0.45\textwidth]{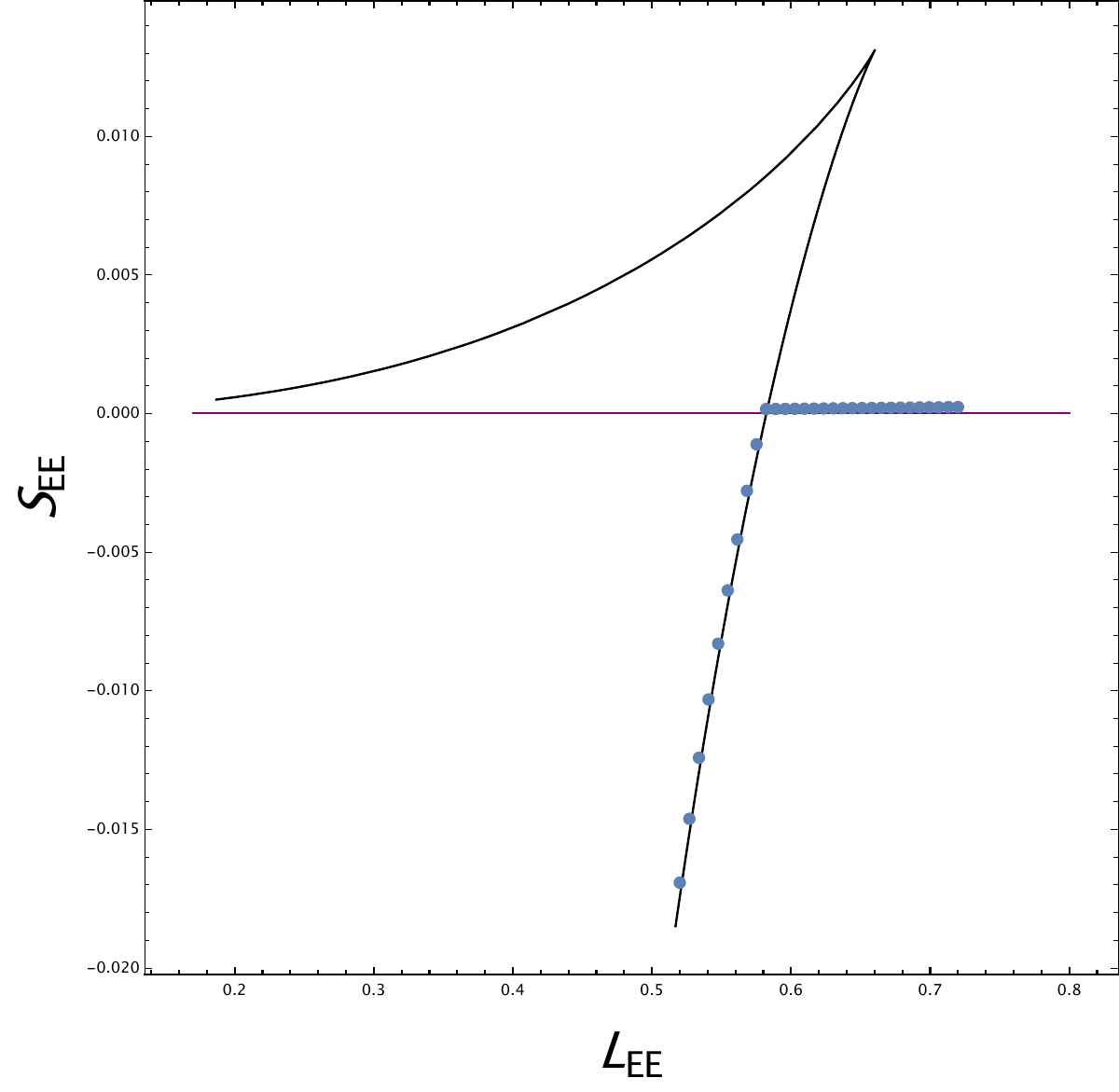}
    \end{center}
    \caption{\small Comparison of results obtained from numerical integration and numerical optimization for case I, presented in \ref{sec:rofx}. Compared to figure \ref{fig:rx_case1}, we gave as an initial guess of the numerical optimization a configuration close to the disconnected one. In this way, the ``stable" branch of the entanglement entropy (right) is always selected, and the vertical line representing the disconnected configuration appears for $L_\text{EE}>L_\text{EE}^\text{crit,2}\simeq0.58$.}
\label{fig:rx_case2}
\end{figure}

Although this case is easily solvable using numerical integration, we have used it as a Toy Problem to validate the numerical strategy used for the more general case presented in section~\ref{sec:genericem}. Knowing that the solution of the Euler-Lagrange equations $r(x_1)$ is a minimum of the function $S_\text{EE}[r(x_1)]$ \eqref{regularized_rx_action}, we compute it via numerical minimization of the action, in a similar manner to the one presented in section~\ref{sec-numerical} or in \cite{Giliberti:2024eii}. The results are in very good agreement with the analytical calculation, as it is visible in figures \ref{fig:rx_case1}, \ref{fig:rx_case2}. Especially, there is a phase transition happening when the disconnected configuration takes over for $L_\text{EE}>L_\text{EE}^\text{crit,2}\simeq0.58$.

\subsection{Case II: \texorpdfstring{$r(z)$}{r(z)}} \label{sec:rofz}

Let us consider the peculiar case where the radial coordinate only depends on the ``quiver direction" $r=r(z)$. The calculation follows in a similar way as the previous embedding, with the determinant of the induced metric now being:

\begin{equation}
    \mathrm{det(g_{\gamma_8})}=-\frac{279936\pi^8\alpha^5\sin^2\theta}{l^6(v_1^2+v_2^2-1)^4\ddot{\alpha}(2\dot{\alpha}^2-3\alpha\ddot{\alpha})^2}\Big[ r^6f(r)\ddot{\alpha} - 6l^2\alpha  (\partial_{z}r)^2\Big],
\end{equation}
while the action is 
\begin{equation}
\begin{split}
&S_{\mathrm{EE}}={\cal N}_{z}\int _{0}^{z_{\star}}\mathrm{d}z\sqrt{F_{z}^2+G_z^2(\partial_{z}r)^2},\\
&{\cal N}_{z}=\frac{L_{x_1}L_{x_2}L_{\phi}\mathrm{Vol}(\Sigma)}{486 \pi G_N},\quad G^2_{z}(r,z)=-\frac{6r^4\alpha^3\ddot{\alpha}}{l^4},\quad F_z^2(r,z) =\frac{f(r)r^6\alpha^2\ddot{\alpha}^2}{l^6}.
\end{split}
\end{equation}

\begin{figure}[htp]
    \begin{center}
    	\includegraphics[width=0.9\textwidth]{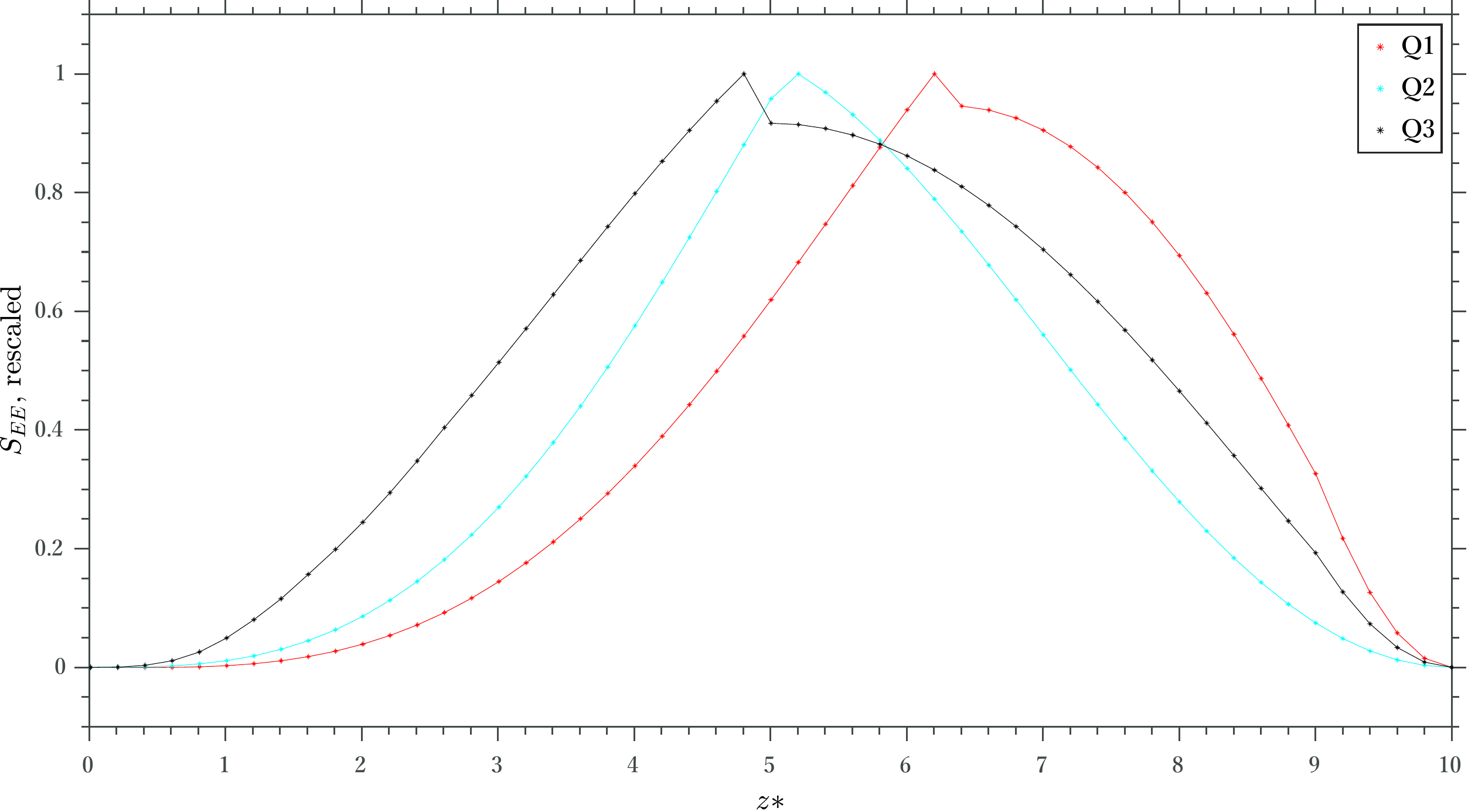}
    \end{center}
    \caption{\small Numerical results for case II, presented in \ref{sec:rofz}, for the three different quivers. On the vertical axis, we show the EE rescaled by their maximum value to highlight the qualitative difference among the quivers.}
\label{fig:rz}
\end{figure}

We numerically solve this using the same compactification and regularization mentioned in sections \ref{sec:rofx}, \ref{sec:genericem}, i.e. finding the configuration $r(z)$ which minimizes the regularized action ($r_{\text{max}}$ denoting the numerical cutoff):
\begin{equation}
    S_{\mathrm{EE}}={\cal N}_{z}\Bigg[\int _{0}^{z_{\star}}\mathrm{d}z\sqrt{F_{z}^2+G_z^2(\partial_{z}r)^2}\,
    -\int_{r_*}^{r_\text{max}}\mathrm{d}r\left(\,G_{{z}}\Big |_{ {z}=0}+\,G_{{z}}\Big |_{ z=z_\star}\right)\Bigg],
\end{equation}
for different values of the parameter $z_\star$. The results, visible in figure \ref{fig:rz}, show how for the three different quivers the EE changes with respect to the length $z_\star$. 

The interpretation of this embedding is that one studies the entanglement not between the typical spatial regions of the $\mathrm{QFT}$, but rather between different parts of the quiver. We effectively consider splitting the quiver into two parts $[0,z_{\star}]\cup (z_{\star},P]$ where each of them contains certain gauge groups (for example in the quiver of figure \ref{quiver_fig} this could be $SU(N_1)\times\dots\times SU(N_\star)$ and $SU(N_{\star}+1)\times\dots\times SU(N_{P-1})$ where $N_{\star}={\cal R}(z_{\star})$) and study the entanglement of the degrees of freedom encoded in $[0,z_{\star}]$ with its complement. 

The resulting EE curve in figure \ref{fig:rz} as a function of division parameter $z_\star$ has a suggestive form. For instance, the $S_\text{EE}$ for the quiver I is initially increasing. This captures the fact that more mutual information is created between the two subdivisions of the quiver for smaller values of $z_\star$. Gradually, the entanglement reaches a maximum and decreases to zero at $z_\star=P$ as the whole range of the quiver is included and the complement subsystem is of zero size. The maximum occurs in a position $z_\star$ leaning towards the gauge node which carries flavor degrees of freedom, showing the effect of these degrees of freedom in the quiver. For the cases of quiver II and III, the relative shape of the quiver is affecting the $S_\text{EE}$ result and again making the role of internal degrees of freedom non-trivial. 

\subsection{Case III: \texorpdfstring{$r(x_1,z)$}{r(x₁,z)}} \label{sec:genericem}

For the most interesting case, which is the one we focus on in this work, our ansatz is $r=r(x_1,z)$ and it gives the determinant of the induced metric:

\begin{equation}
    \mathrm{det(g_{\gamma_8})}=-\frac{279936\pi^8\alpha^5\sin^2\theta}{l^6(v_1^2+v_2^2-1)^4\ddot{\alpha}(2\dot{\alpha}^2-3\alpha\ddot{\alpha})^2}\Big[ r^6f(r)\ddot{\alpha}-6l^2\alpha r^2 (\partial_z r)^2 + l^4\ddot{\alpha} r^2 (\partial_{x_1}r)^2\Big],
\end{equation}
leading to the following action, now including an integration up to a point $z_\star$ in the $z$ direction as well:
\begin{equation}
\begin{split}
    S_{\text{EE}}
    &={\cal N}_{x_1,z}\int_{-L/2}^{L/2}\mathrm{d}x_1\int_0^{z_{\star}}\mathrm{d}z\sqrt{G_{x_1} ^2(\partial_{x_1}r)^2+G_z^2(\partial_z r)^2 + F^2},\\
    &G_{x_1}^2(x_1,z)=l^4r^2(\alpha\ddot{\alpha})^2,\quad F^2(x_1,z)=r^6f(r)(\alpha\ddot{\alpha})^2,\quad G_z^2(x_1,z)= -6l^2r^4\alpha^3\ddot{\alpha},\\
        &{\cal N}_{x_1,z}=\frac{L_{x_2}L_{\phi}\mathrm{Vol}(\Sigma)}{486 l^3\pi G_{N}}.
\end{split} \label{action_full}
\end{equation}

Similarly to the embedding of case II, by allowing the profile of the surface to explore the $z$ coordinate as well as $x_1$, we get the chance to study not only the entanglement between spatial regions in the dual $\mathrm{CFT}$, but also the entanglement of degrees of freedom belonging to different gauge groups in the quiver. 

As mentioned in the introduction, this embedding introduces new technical challenges as one is now dealing with partial differential equations of motion instead of ODEs, which leaves no room for analytic approaches. Instead, we solve the dynamics via a numerical variational approach: in the following section~\ref{sec-numerical} we provide a numerical analysis of the problem.

\FloatBarrier
\section{Numerical approach and results}\label{sec-numerical}

In order to tackle the problem in section~\ref{sec:genericem} via minimization of the action \eqref{action_full}, we employ a numerical optimization algorithm written in \textsc{Matlab} \cite{MATLAB2023}.
Here we show the procedure followed to find the solutions, the outline of the algorithm, the solutions found for three different rank functions, and our analysis of these results. The algorithm works in a similar way to \cite{robinhood}, which was written in the \texttt{Julia} programming language for 1D optimization. 

Before minimizing the action \eqref{action_full} as explained in the next section, we perform some adjustments for better visualization and a faster, more robust resolution.
First, we rewrite the action:
\begin{equation}
    S_{\text{EE}}={\cal N}_{x_1,z} \,\hat{S}_{\text{EE}},\nonumber
\end{equation}
\begin{equation}
    \hat{S}_{\text{EE}}[r(x_1,z)] = \int_{-L/2}^{L/2}\mathrm{d}x_1\int_0^{z_{\star}}\mathrm{d}z\sqrt{G_{x_1} ^2(\partial_{x_1}r)^2+G_z^2(\partial_z r)^2 + F^2}.\label{action-factored}
\end{equation}

The boundary conditions that we impose on the solutions for this action are
\begin{eqnarray}\label{BCs-1}
    r\left(x_1=\pm\frac{L}{2},z\right)=r\left(x_1,z=0\right)=r\left(x_1,z=z_\star\right)=\infty.
\end{eqnarray}
This ensures, as prescribed by \eqref{S_D}, that the boundary of the minimal surface solution in the gravity background is the same as the one of the CFT subsystem.

Then, to have a square integration domain,\footnote{Integration is performed by numerical quadrature on a triangulated domain; rescaling to a square produces more regular elements, which improves computational stability and conditioning of the approximation.} we compactify the integration directions, making a change of variables:
\begin{align}\begin{cases}
    x_1&\rightarrow \quad\hat{x}=x_1/L,\\
    z&\rightarrow \quad\hat{z}=z/z_\star.\end{cases}
\end{align} 
Using the compactification above, the action \eqref{action-factored} can be written as:
\begin{eqnarray}
    \hat{S}_{\mathrm{EE}}= L z_{\star}\int_{-1/2}^{1/2}\mathrm{d}\hat{x}\int_0^1 \mathrm{d}\hat{z}\sqrt{G^2_{\hat{x}}\frac{(\partial_{\hat{x}}r)^2}{L^2} + G^2_{\hat{z}}\frac{(\partial_{\hat{z}}r)^2}{z_{\star}^2}+\hat{F}^2},\label{action_final}
\end{eqnarray}
where now, 
\begin{eqnarray}
    G^2_{\hat{x}}:=l^4r^2(\hat{\alpha}\ddot{\hat{\alpha}})^2,\quad G^2_{\hat{z}}:=- 6l^2r^4\hat{\alpha}^3\ddot{\hat{\alpha}},\quad \hat{F}^2:= r^6f(r)(\hat{\alpha}\ddot{\hat{\alpha}})^2,
\end{eqnarray}
and $\hat\alpha$ is the scaled version of the ordinary $\alpha$, meaning
\begin{equation}
    \hat\alpha=\alpha(z_{\star}\hat{z}).
\end{equation}
Inside the scaled $\alpha$, the parameter $P$ is still present. We will use the three rank functions mentioned in section~\ref{section-quiver-QFT}, rescaled to have actions of $\mathcal{O}(1)$ for a more straightforward setup of the optimization parameters.

Moreover, in the $r$-direction, we implement a cutoff at $r=r_\text{max}$. Thus, the boundary conditions \eqref{BCs-1} become 
\begin{eqnarray}\label{BCs-2}
    r\left(\hat{x}=\pm\frac{1}{2},\hat{z}\right)\,=\,r\left(\hat{x},\hat{z}=0\right)\,=\,r\left(\hat{x},\hat{z}=1\right)\,=\,r_\text{max}; \qquad \quad r\in(r_*,r_\text{max}].
\end{eqnarray}

Following sections \ref{sec:rofx} and \ref{sec:rofz}, we regularize by subtracting the action of ``trivial solution", i.e. the action for the disconnected configuration\footnote{We use the term \textit{disconnected} for continuity with the previous sections, even though the configuration is connected, albeit not simply connected.} extending from the boundary at $r=r_\text{max}$ down to $r=r_*$:
\begin{equation}
\begin{split}
    \hat{S}_{\mathrm{EE},0}
    &= Lz_\star\Bigg[ \int_{r_*}^{r_\text{max}}\mathrm{d}r\int_0^1\mathrm{d}\hat z\,G_{\hat x} \Big |_{\hat x=-1/2}+\int_{r_*}^{r_\text{max}}\mathrm{d}r\int_0^1\mathrm{d}\hat z\,G_{\hat x} \Big |_{\hat x=1/2}\\
    &+\int_{-1/2}^{1/2}\mathrm{d}\hat x\int_{r_*}^{r_\text{max}}\mathrm{d}r\,G_{\hat{z}}\Big |_{\hat {z}=0}+\int_{-1/2}^{1/2}\mathrm{d}\hat x\int_{r_*}^{r_\text{max}}\mathrm{d}r\,G_{\hat{z}}\Big |_{\hat z=1}\Bigg].
\end{split}\label{regulatization}
\end{equation}

For simplicity of reading, from here on we drop the hat $\hat{*}$ from the compactified variables, as we will always use their compactified adimensional versions.

\subsection{Algorithm}\label{sec:algor}

Preliminarily, we recapitulate what our algorithm needs to do: we seek the solution $r(x,z)$ that minimizes the integral action \eqref{action_final} regularized via \eqref{regulatization}; we will evaluate the action on this \textit{optimal} (i.e. minimal) \textit{configuration} and study its dependence on the free parameters $L$ and $z_\star$, as well as on different rank functions of the boundary quiver theory; we will also study how the turning point\footnote{For a symmetric one-dimensional case as section~\ref{sec:rofx}, the definition of $r_0$ as the turning point is straightforward. We are aware that it is not for this more general case; however, we still choose to define $r_0$ as the minimum value that the solution has across the integration domain.} $r_0$ of the minimal configuration depends on the same parameters. Since the computations for different values of $L$ and $z_\star$ are all independent from each other, we employ (single-node, multicore) parallel programming.

We treat the integration in \eqref{action_final} as a 2D integration on a domain $\Omega_{x,z}\coloneqq[-1/2,1/2]\times[0,1]$:
\begin{equation}
    \int_{-1/2}^{1/2}\mathrm{d}x\int_0^1 \mathrm{d}z = \iint_{\Omega_{x,z}}\dd A.
\end{equation}
We define a set of triangles $\bigtriangleup\coloneqq\{T_1\dots T_N\}$ to be a triangulation of the domain $\Omega_{x,z}=\bigcup_{i=1}^N T_i$ if two distinct triangles with non-empty intersection either share a single vertex or a whole common edge. The triangulation will have $n_v$ vertices and $n_t$ triangles, and will depend on two parameters $(h,H)$ such that
\begin{equation}\label{h}
    h\coloneqq\max_{T\in\bigtriangleup}\abs{T}, \qquad h/H\coloneqq\max_{T\in\partial\bigtriangleup}\abs{T}.
\end{equation}
Using two different values for the maximum size of the triangles' sides on the interior and on the boundary gives the possibility of having more resolution where it is needed, while keeping the total number of triangles low.

We approximate the function $r(x,z)$ via the use of splines \cite{splines,splinesontrg}, i.e. globally regular piecewise polynomial functions, on the triangulated domain $\Omega_{x,z}$; we use splines of degree $(1,0)$: globally continuous affine functions on every triangle of the triangulation. In practice, the spline $s[x,z]$ is stored as its \textit{B-form}: an array of length $n_v$, with the $i$-th element of the array approximating the value of $r(x,z)$ at the coordinates of the $i$-th vertex of $\bigtriangleup$. The B-form of a spline can be written as 
\begin{equation}\label{bform}
    s=\sum_{i=1}^{n_v}c_i\varphi_i,
\end{equation}
where $c_i$ are the coefficients and $\varphi_i$ are the triangulation's \textit{hat functions}, i.e. splines that are zero at every vertex except for the $i$-th, which serve as a basis.

The action $S_\text{EE}$ is a function of the spline $s$ and the parameters $(L,z_\star)$ that returns a real value. It is calculated using midpoint composite quadrature integrating the Lagrangian. Moreover, for a faster resolution, we also write a function for the gradient of the
action with respect to the B-form coefficients $c_i$: 
\begin{equation}
    \begin{split}
        S_\text{EE}[s(x,z)]&=\iint_{\Omega_{x,z}}\mathcal{L}(s,\partial_x s,\partial_z s)\,\dd A \\
        &\Rightarrow\; \nabla S_\text{EE}[s(x,z)]_i=\iint_{\Omega_{x,z}}\frac{\partial\mathcal{L}(s,\partial_x s,\partial_z s)}{\partial c_i}\,\dd A,\\
        &\frac{\partial\mathcal{L}(s,\partial_x s,\partial_z s)}{\partial c_i}=\frac{\partial \mathcal{L}}{\partial s_b}\frac{\partial s_b}{\partial c_i}+\frac{\partial \mathcal{L}}{\partial (\partial_x s)}\frac{\partial (\partial_x s)}{\partial c_i}+\frac{\partial \mathcal{L}}{\partial (\partial_z s)}\frac{\partial (\partial_z s)}{\partial c_i},
    \end{split}
\end{equation}
where $s_b$ is the value of the spline evaluated at the barycenter of the triangles, and the partial derivatives $\partial/\partial c_i$ are computed as derivatives of the hat functions. The gradient thus returns $n_v$ real values.

The action \eqref{action_final} together with its gradient, the constraints and the bounds \eqref{BCs-2}, is passed to an optimization solver that uses Sequential Quadratic Programming \cite{NoceWrig}, an iterative method for constrained nonlinear optimization.\footnote{The reason for this choice is that SQP methods satisfy bounds at each iteration, they use a merit function that combines the objective and constraint functions, and they are faster than interior-point Newton methods on small-to-medium scale problems, such as ours.} It is a local optimization solver, therefore we test the well-posedness of our problem (for which we seek a global minimum) by providing multiple initial guesses $s_0$. With a-posteriori analysis we notice that, similarly with what is shown in figures \ref{fig:rx_case1} and \ref{fig:rx_case2} for the 1D case, for most values of $(L,z_\star)$ the optimal solution is indeed independent on the initial guess, while for some intermediate values there are two competing configurations which are selected depending on the initial guess: one close to the disconnected configuration, and a smooth connected one. In order to properly treat these cases, we perform each optimization twice, with two different initial guesses, and only the result that yields the lowest action is recorded.
Finally, the optimal value of $S_\text{EE}$, together with the parameters used to produce it and the optimal spline $s$, is stored to be analyzed.

Therefore, specifying the desired $\alpha(z)$ function, the parameters of the theory $(l,\mu,q,P)$, and the lengths $(L,z_\star)$, our algorithm can find the configuration that solves the corresponding equations of motion and find the EE of such a configuration. Of all these variables, we explore how different $\alpha(z)$ behave when changing $(L,P,z_\star)$, fixing $\mu=0$ (to be in the simpler SUSY case) and $q=1,\,l=1$. 
\\
Below, we present the results in the case of three different quivers. We follow the same logic and order in the presentation of each of them, with emphasis on the physical meaning of our results.

\subsection{Solutions}

Using the algorithm presented in \ref{sec:algor}, we compute the entanglement entropy and the entanglement surface for a variety of cases. We chose three different functions $\alpha(z)$ resulting in three different rank functions $R(z)$. For each of them, we run the algorithm for multiple values of $(L,P,z_\star)$ and record, for each case, the function $r(x,z)$ that minimizes the action and the value of the minimized action, i.e. the entanglement entropy $S_\text{EE}$. Here we present the data and the solutions found.
The results shown are obtained, except when otherwise specified, for $r_\text{max}=30$, $P=10$, $h=0.24$, $H=12$, and choosing the quiver I (scalene triangle).

\begin{figure}[htp]
    \begin{center}
        \includegraphics[width=0.49\textwidth]{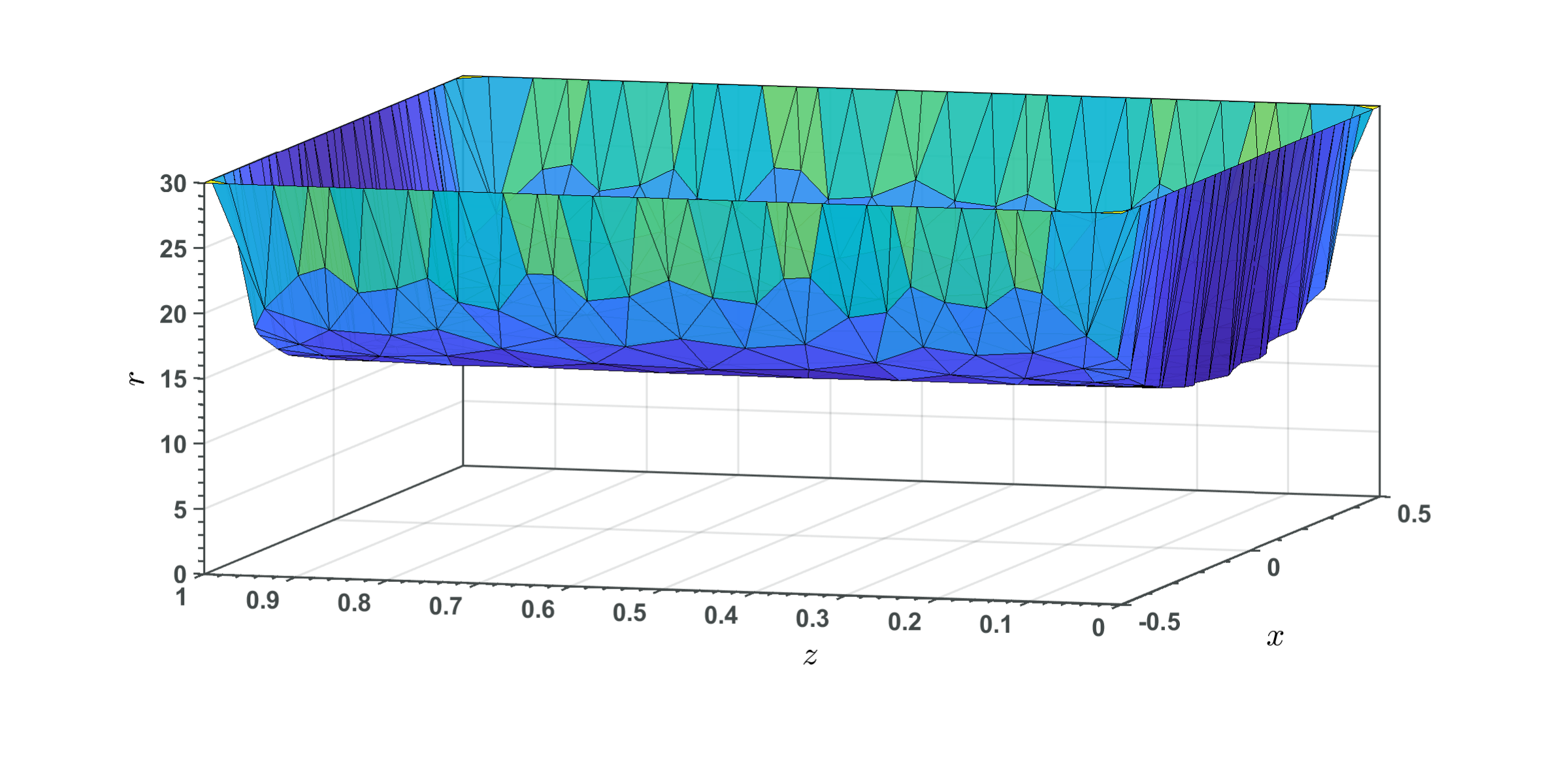}
        \includegraphics[width=0.49\textwidth]{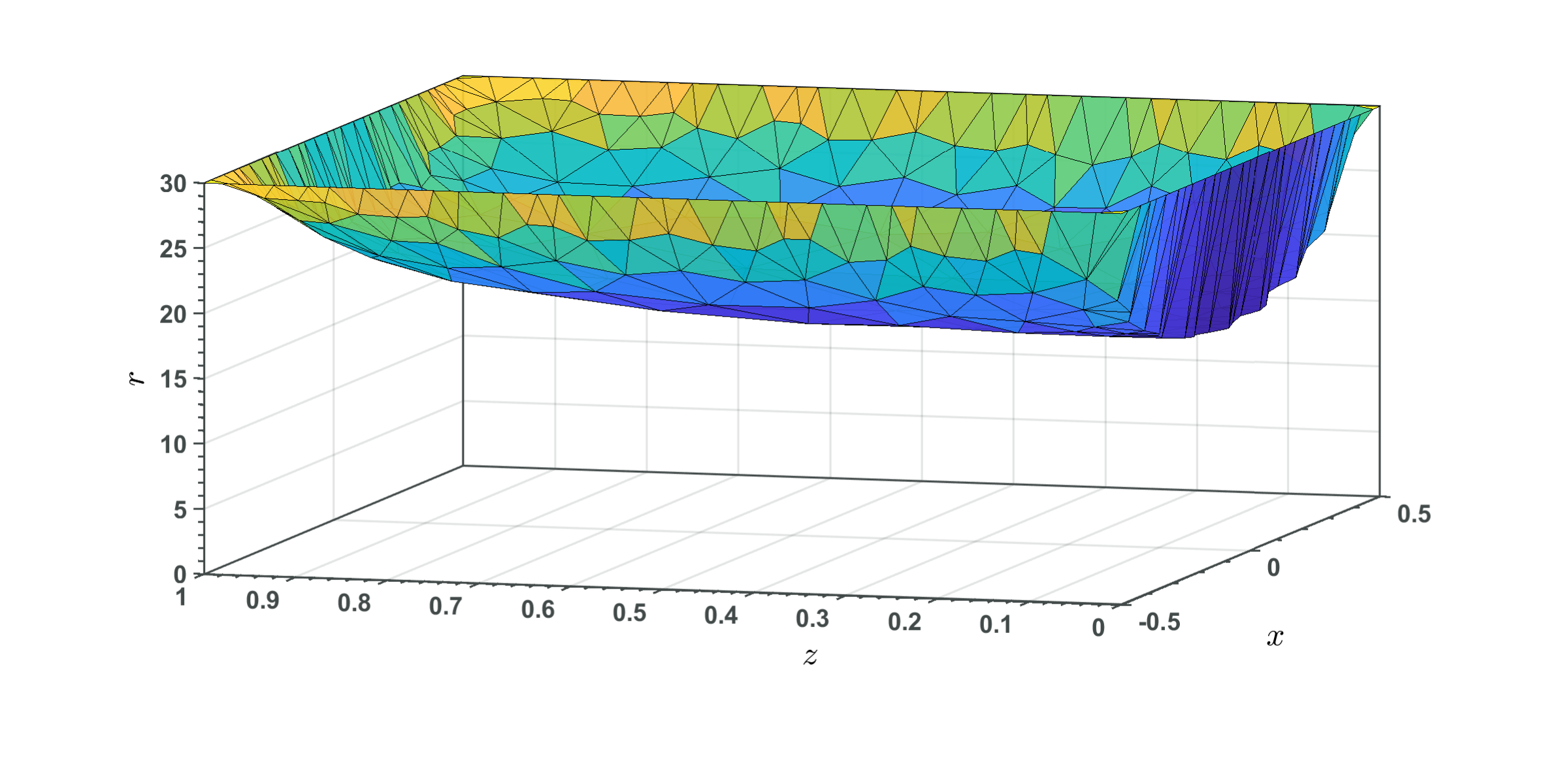}
        \includegraphics[width=0.49\textwidth]{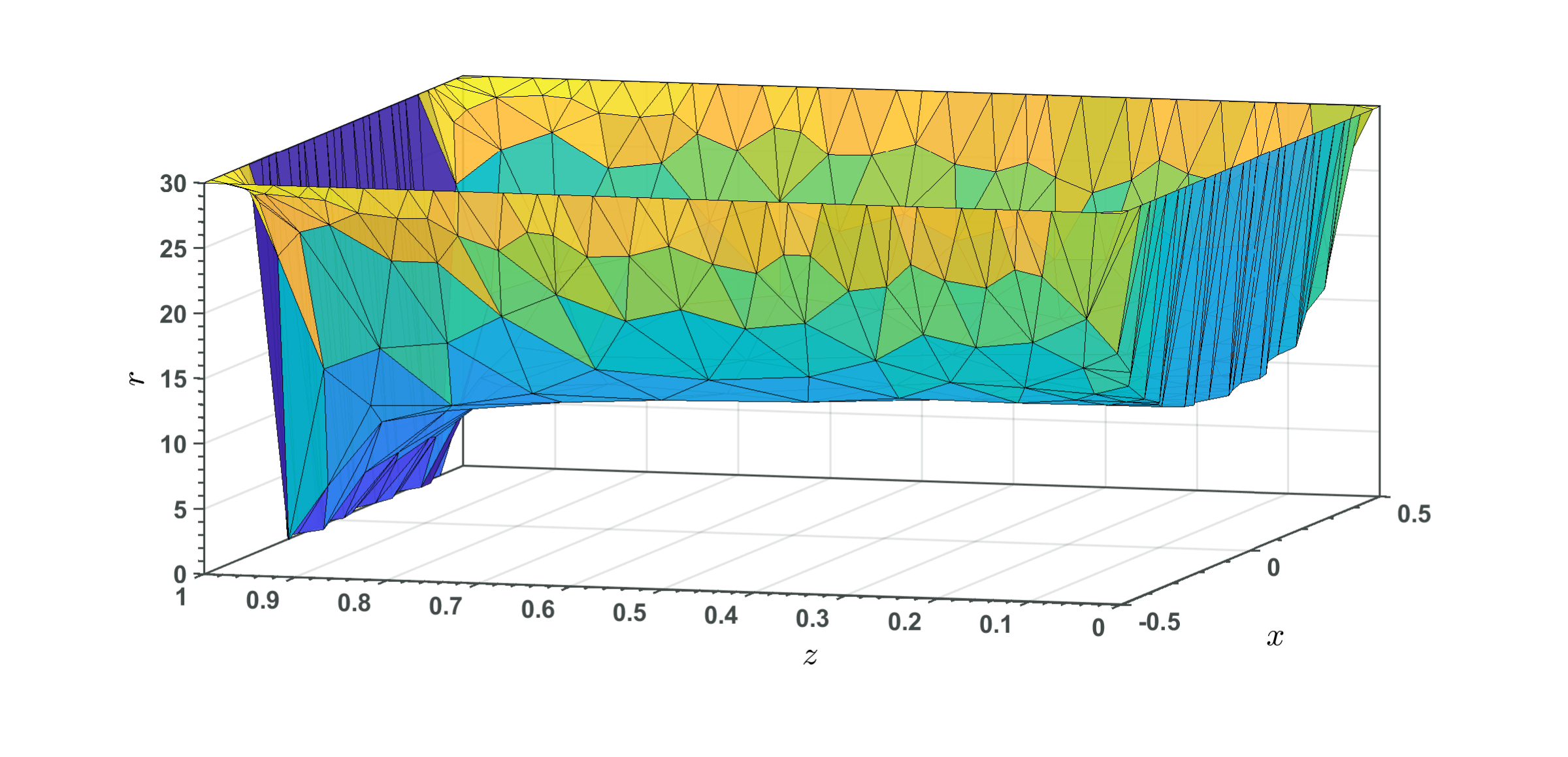}
        \includegraphics[width=0.49\textwidth]{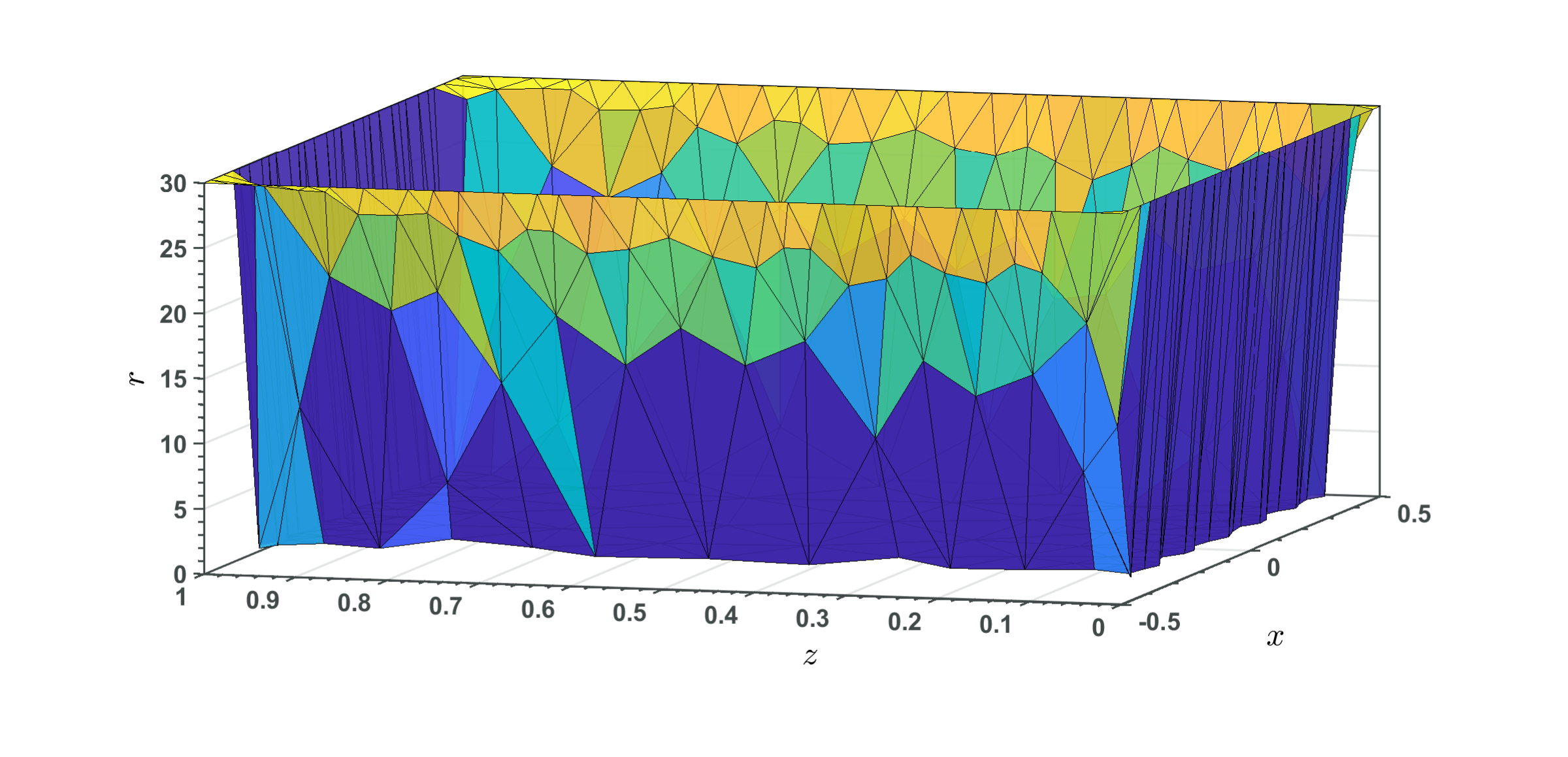}
    \end{center}
    \caption{\small The optimal spline solution $r(x,z)$ for various values of the parameter $z_\star$: $z_\star=10$ (top-left); intermediate values $z_\star\sim8$ and $z_\star\sim6$ (top-right, bottom-left); $z_\star\sim0$ (bottom-right). The parameter $L$ interpolates between connected configurations (top-row-like) and disconnected ones (bottom-row-like). It can be seen how for $z_\star\sim6$ (bottom-left) only some values of $z$ deconfine, signaling a partial deconfinement as discussed in section \ref{sec:discussion}.}
\label{fig:rxz_solution_multi}
\end{figure}

In figure \ref{fig:rxz_solution_multi}, examples of multiple optimal solutions for the $r(x,z)$ profile are provided. The profiles have non-trivial $z$ dependence, signaling the effect of the whole quiver degrees of freedom on the calculation of $S_\text{EE}$. Similar observations had been made about the profile of the probe string dual to the Wilson loop expectation values in these setups in ref.~\cite{Giliberti:2024eii}. Indeed, close to the presence of flavor branes, the profile starts to change considerably as in the bottom-left panel of the above-mentioned figure.

\begin{figure}[htp]
    \begin{center}
    	\includegraphics[width=0.9\textwidth]{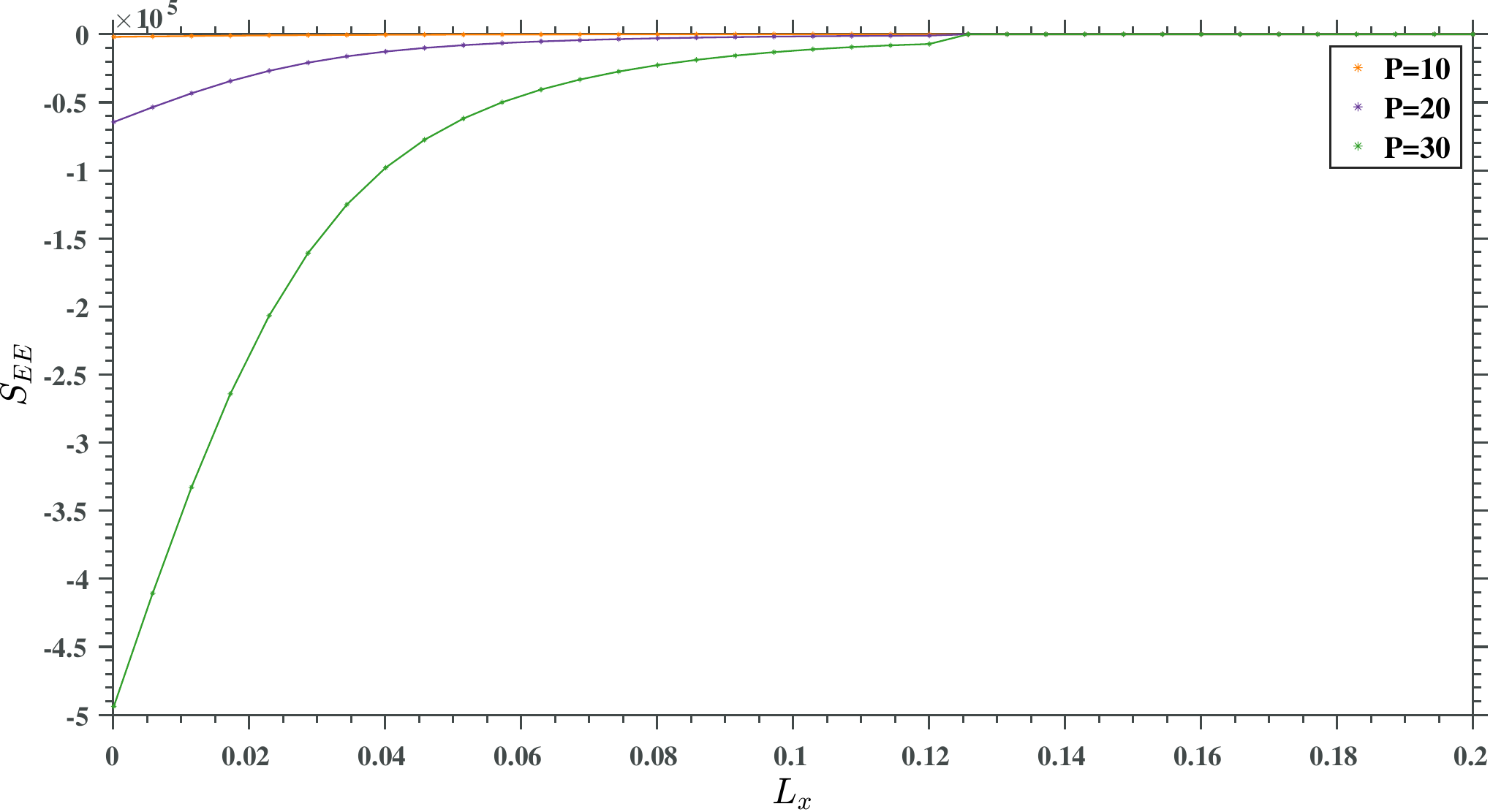}
    \end{center}
    \caption{\small Entanglement entropy as a function of the spatial separation, with $z_\star=P$, for various values of P.}
\label{fig:SofL}
\end{figure}

Figure \ref{fig:SofL} reproduces the results obtained in section~\ref{sec:rofx}, but takes into account the $z$ dependence of the profile. The calculation is done for quiver I and different values of $P$ by varying the $L$ separation. As the profiles provided in figure \ref{fig:rxz_solution_multi} suggest, the dependence on $z$ introduces some changes to the values of $S_\text{EE}$, although the schematic behavior is the same as in section~\ref{sec:rofx}. Since our numerical method is designed to detect the true minima of the action,\footnote{As previously mentioned, we highlight that we employ a local optimizer; the true minima are found by performing various initial guesses and comparing the outcomes, effectively implementing a multi-start strategy that mimics global optimization.} only the physically preferred branch of the solution is obtained. Hence, the metastable (and obviously the unstable) branches present in figure \ref{fig:rx_case2} are absent in these results. Nevertheless, the discontinuity in the slope of $S_\text{EE}$ is visible in the plots, signaling the phase transition to the trivial solution after certain separation values $L$.

\begin{figure}[htp]
    \begin{center}
    	\includegraphics[width=0.9\textwidth]{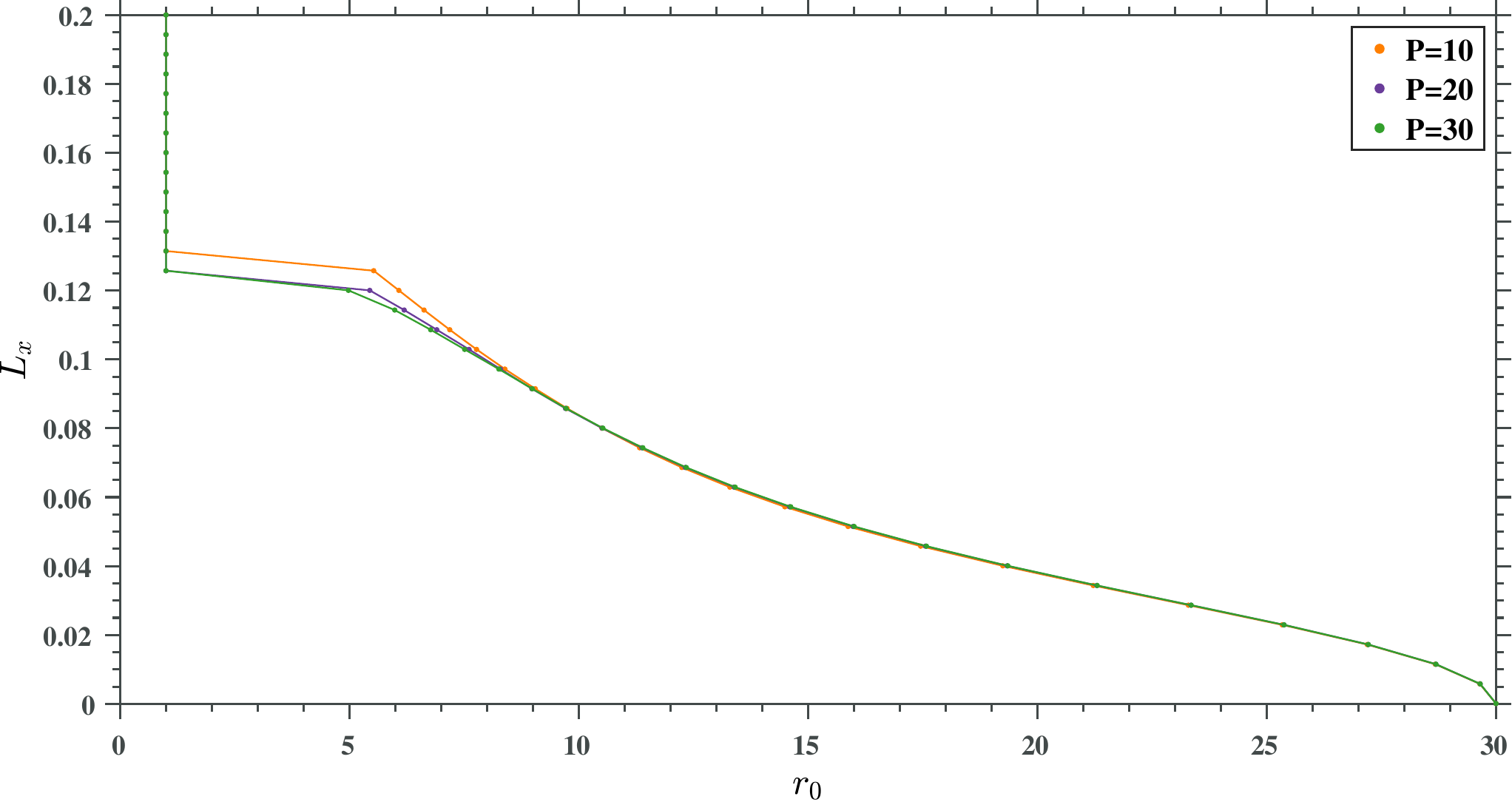}
    \end{center}
    \caption{\small The spatial separation $L$ as function of the turning point $r_0$, for various values of the quiver size $P$, for $z_\star=P$. The phase transition is visible, and the shape recalls the result from the 1D case in figure \ref{fig:rx_case2}. The results for the other quivers are qualitatively similar.}
\label{fig:Lofr0}
\end{figure}

\begin{figure}[htp]
    \begin{center}
    	\includegraphics[width=0.9\textwidth]{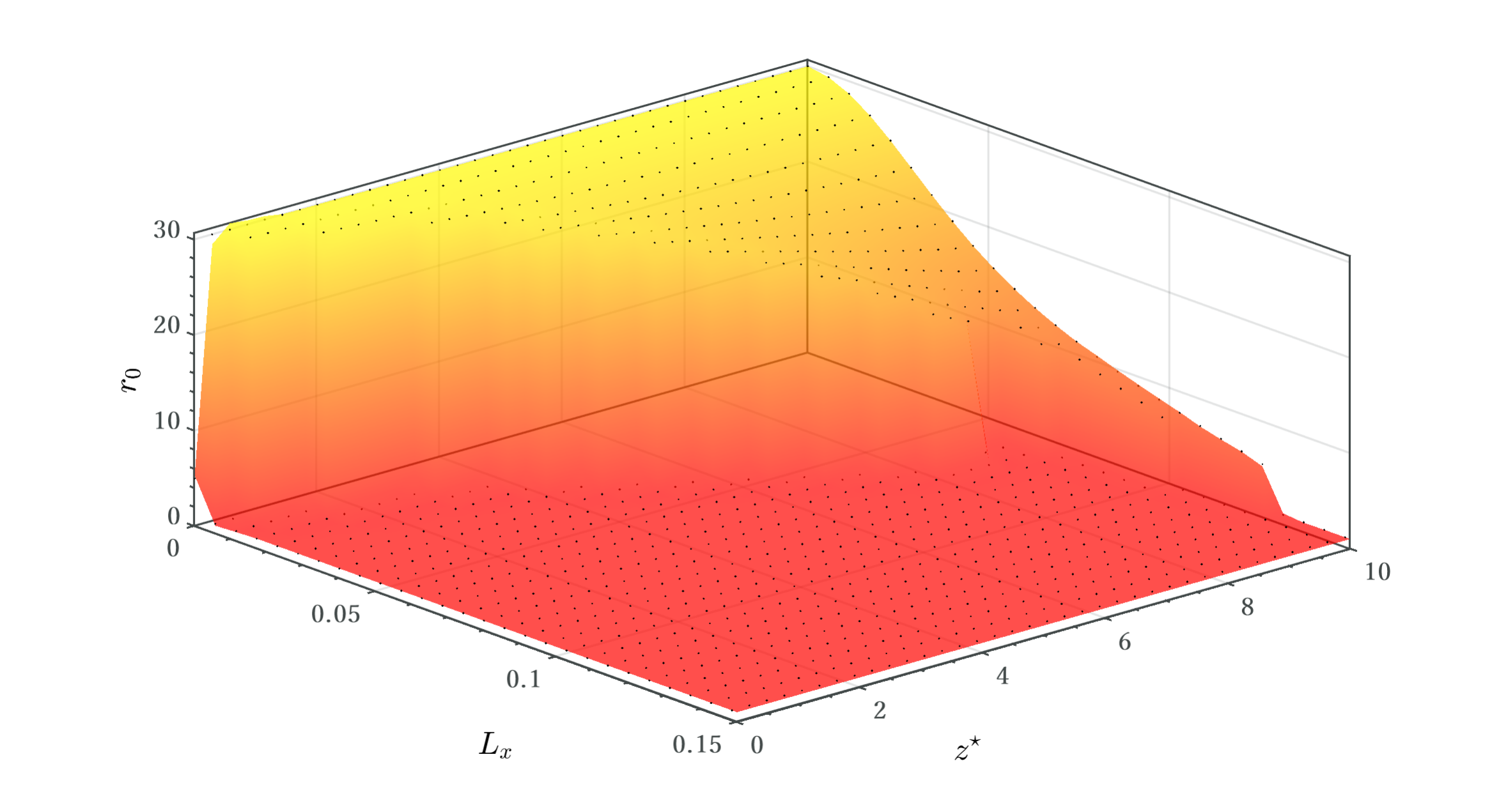}
    \end{center}
    \caption{\small The turning point $r_0$ as function of the lengths $(L,z_\star)$. The phase transition is clearly visible, with two distinct regions (``connected" for $r_0\neq 1$, ``disconnected" for $r_0=1$). The results for the other quivers are qualitatively similar.}
\label{fig:r0_Lzs}
\end{figure}

The dependence of $L$ on the turning point parameter $r_0$ for quiver I is depicted in figure \ref{fig:Lofr0}. The phase transition by jump in the value of $r_0$ at large $L$ separations is present in this plot and it can be related to confining-like behavior, similarly to figure \ref{fig:rx_case2}. 
Figure \ref{fig:r0_Lzs} depicts the complete dependence of $r_0$ on $L$ and $z_\star$, with the previous figure being a slice of this 3D plot. Here, we can see the range of values for $L$ and $z_\star$ where the phase transition occurs, with the two regions clearly separated. We must bear in mind that the definition of $r_0$ in these two plots is not as precise as $r_0$ in \eq{eq:Hr0}: for 3D profiles of $r(x,z)$, the turning point $r_0$ can be defined in different ways, and we choose the minimum value of $r$ in the whole 2D space $(x,z)$ as our definition. 

\begin{figure}[htp]
    \begin{center}
    	\includegraphics[width=0.7\textwidth]{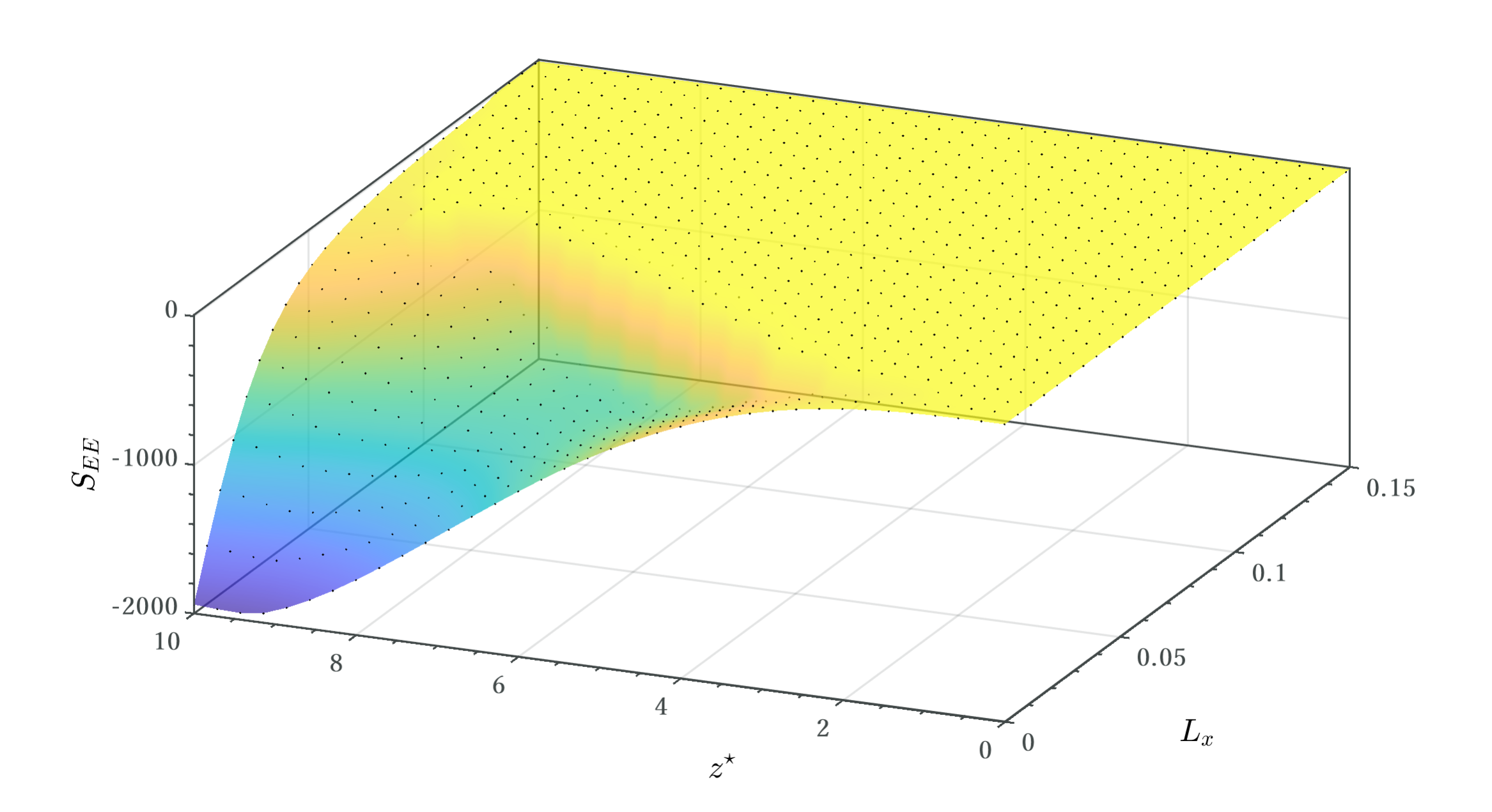}
        \includegraphics[width=0.7\textwidth]{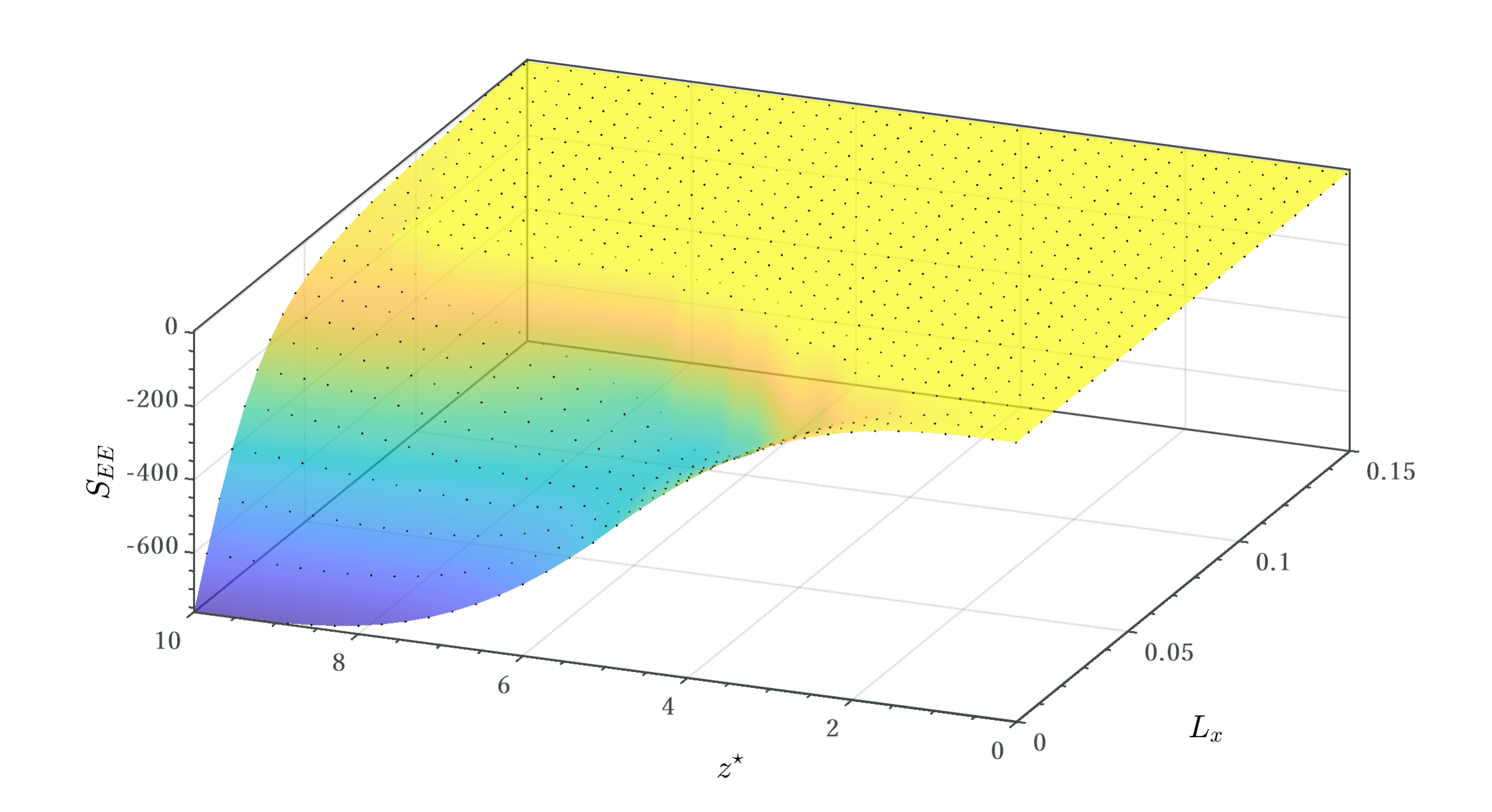}
        \includegraphics[width=0.7\textwidth]{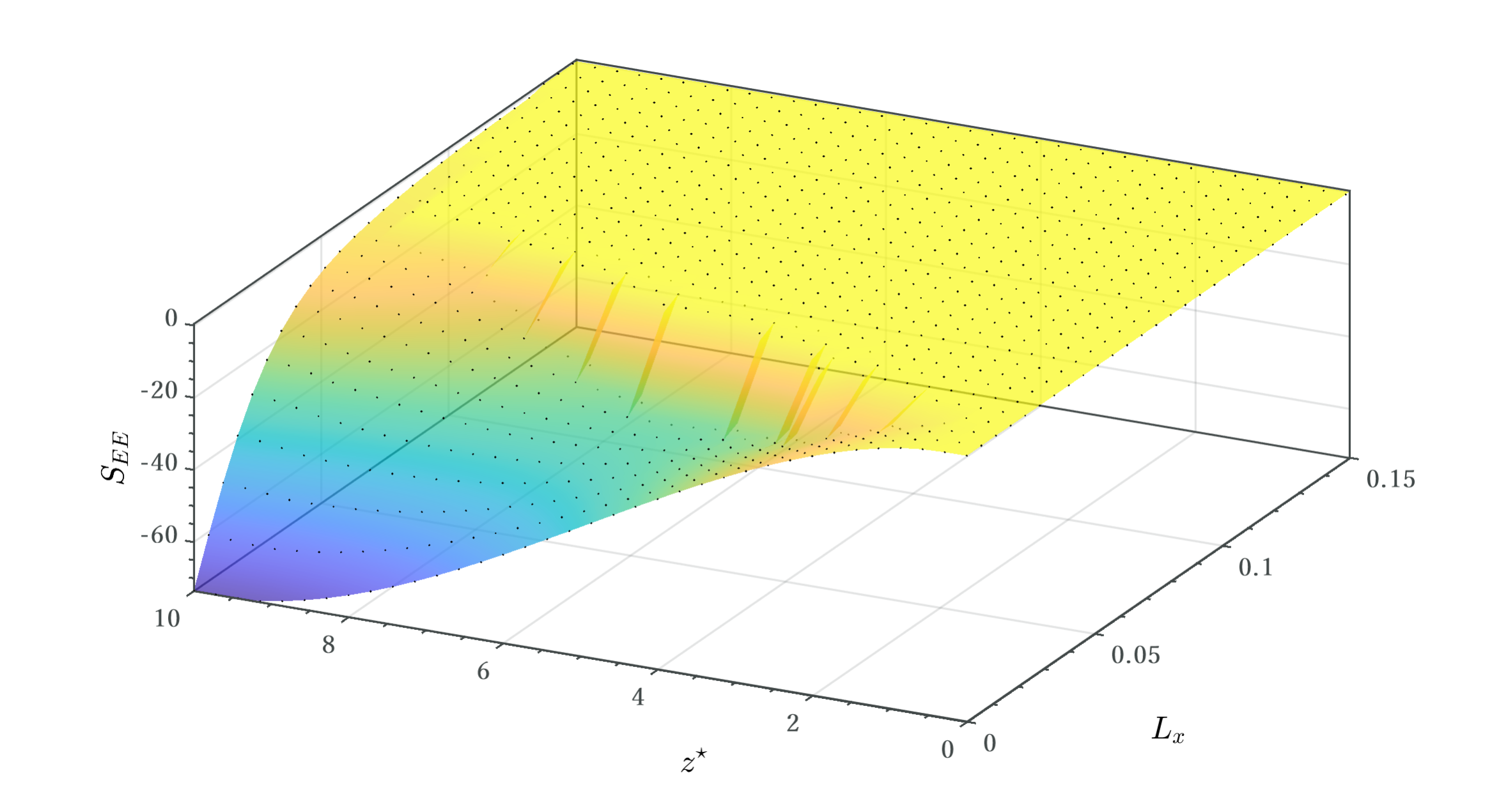}
    \end{center}
    \caption{\small Entanglement entropy as a function of the two lengths $L$ and $z_\star$, for quiver I (top), II (middle), and III (bottom).}
\label{fig:S_Lzs_multiQ}
\end{figure}

Figure \ref{fig:S_Lzs_multiQ} provides the final complete result for $S_\text{EE}$ as a function of $(L,z_\star)$ in quivers I, II, III; the previous $S_\text{EE}$ plots in this section are slices of this diagram. The interesting feature is the appearance of a sharp jump in the values of $S_\text{EE}(L,z_\star)$, for the range $0<z_\star<P$, especially for the middle values of $z_\star$. This can be interpreted as a zeroth-order phase transition, and its theoretical details and explanation need further study. We provide some preliminary interpretation in the next section. 

\begin{figure}[htp]
    \begin{center}
    	\includegraphics[width=0.7\textwidth]{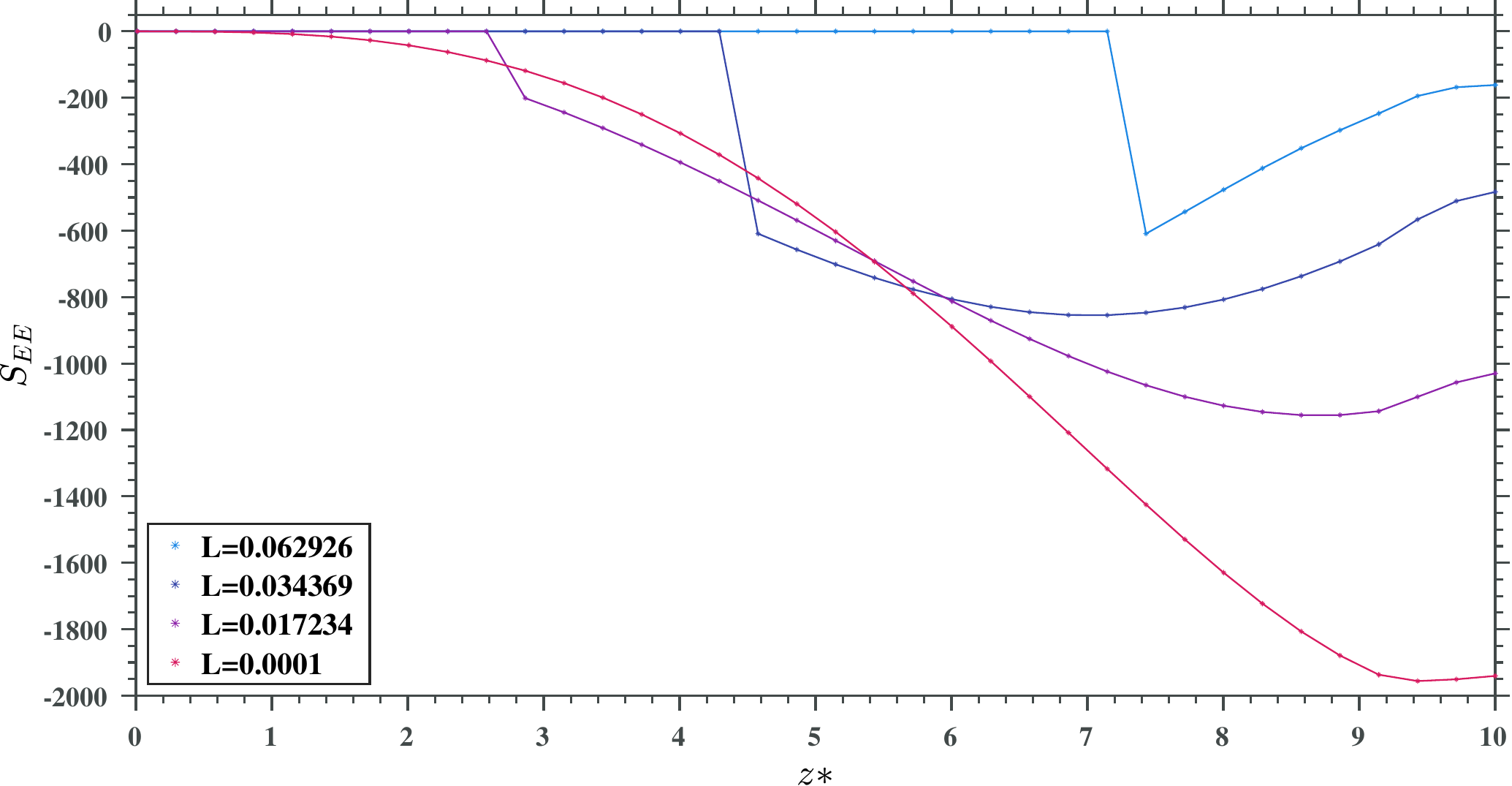}
        \includegraphics[width=0.7\textwidth]{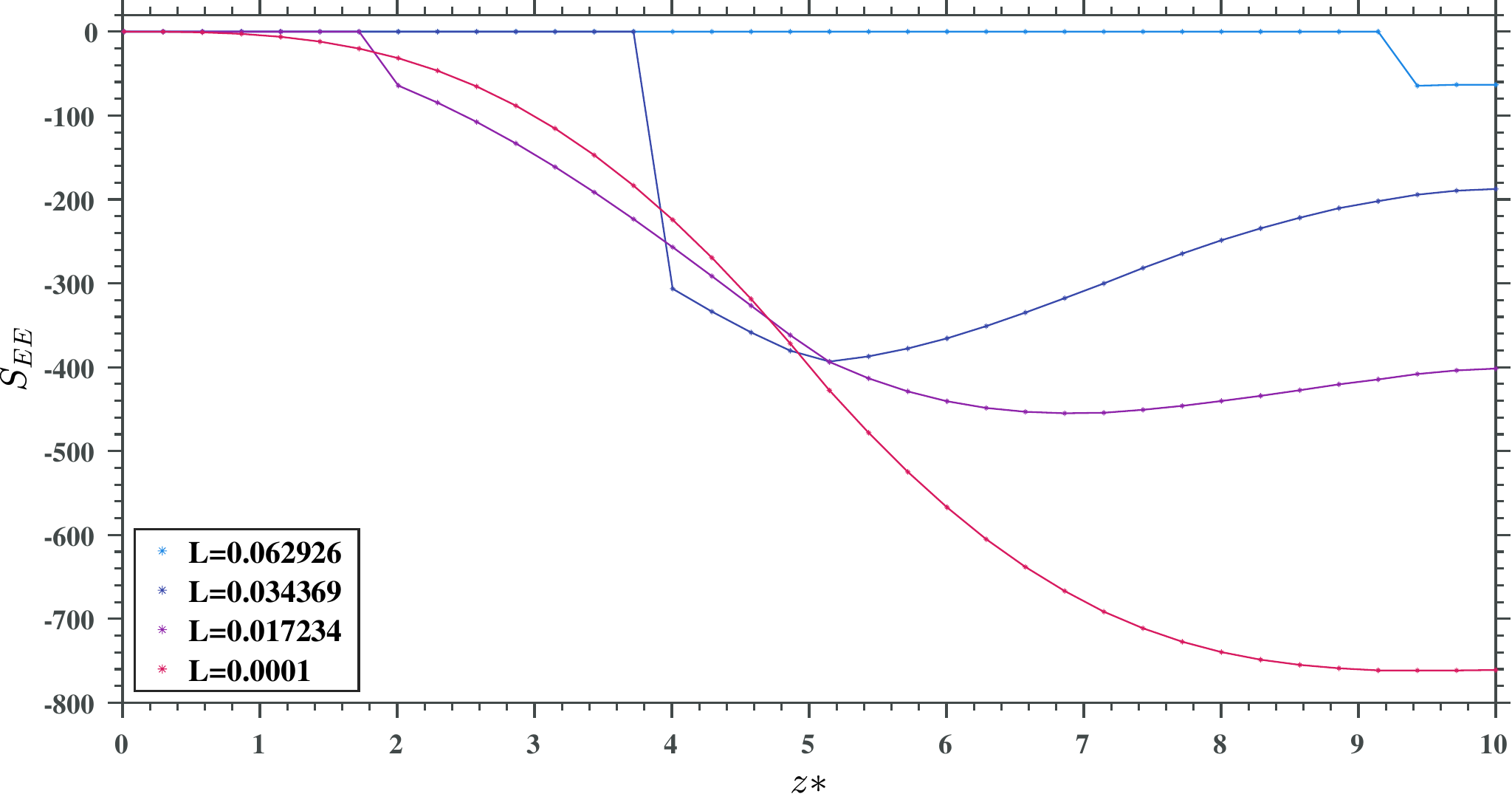}
        \includegraphics[width=0.7\textwidth]{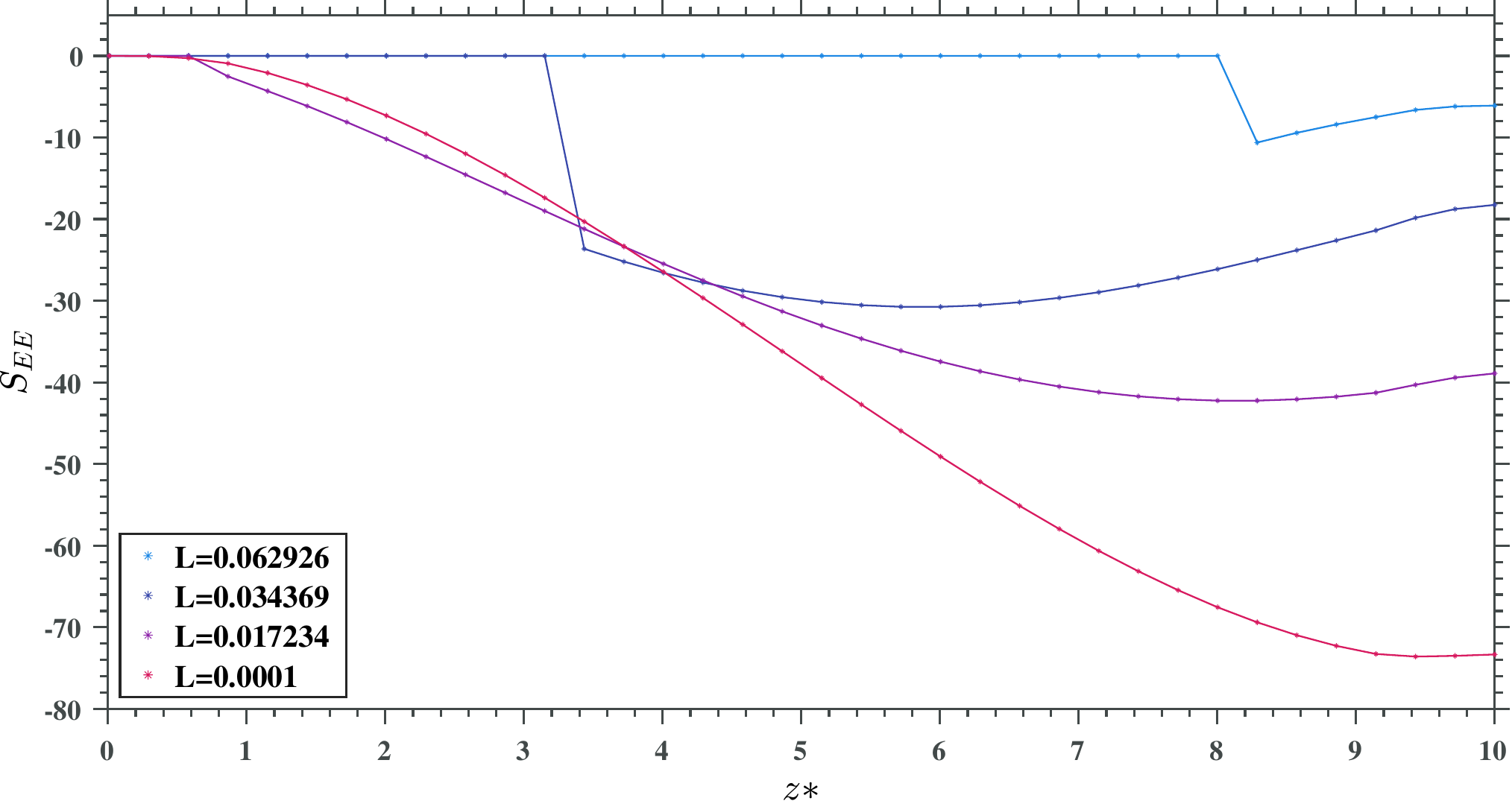}
    \end{center}
    \caption{\small Entanglement entropy as a function of the quiver length $z_\star$ for various values of L, for quiver I (top), II (middle), and III (bottom). The phase transition is clearly visible, and the $(L,z_\star)$ values for which it appears are shown here to be influenced by the quiver choice.}
\label{fig:Sofzs_multiQ_multiL}
\end{figure}

Figure \ref{fig:Sofzs_multiQ_multiL} pinpoints the dependence of $S_\text{EE}$ on $z_\star$ more clearly for quivers I, II, and III for certain chosen values of $L$, and it is intended as a better visualization for figure \ref{fig:S_Lzs_multiQ}. First of all, the plots show the effect of different quiver shapes on the EE value, again showcasing the important effect of internal degrees of freedom. The common trend is a decrease in the values of $S_\text{EE}$ to a minimum, followed by an increase as the range of integration in $z$ increases. The position of the flavor branes affects the shape of the curve. This should be compared with the results for section~\ref{sec:rofz}, as the trend here is altered due to the introduction of an $x$ dependence.

\subsection{Discussion}\label{sec:discussion}

Let us briefly review the results obtained in this work. By implementing numerical methods to minimize the action for the probe surface, we obtained the minimal RT surface for the calculation of EE. As long as the whole range of the quiver is taken into account ($z_\star=P$), there is a qualitative match with the previous studies of simpler cases considering only $S_\text{EE}$ as a function of $x$, though quantitatively the gauge and flavor degrees of freedom have some effects, especially altering the profile of the RT surface. 

More interestingly, in section~\ref{sec:rofz}, we studied the effects of a separation of regions for the quiver degrees of freedom while keeping the space regions unchanged and united. In this case, a clear dependence on gauge/flavor degrees of freedom appears, and it is also sensitive to the shape of the quiver. 

Finally, the whole dependence of $S_\text{EE}$ on $(L,z_\star)$ combinations is considered. In addition to sensitivity to the shapes of the quiver, a peculiar phase transition structure appears in the internal points of the $z$ direction that may help to study the phase diagram of the QFTs under consideration. 

Here, we want to draw the reader's attention to a similar observation in the literature in the context of partial deconfinement. As it is studied and explained in detail in \cite{Hanada:2016pwv, Hanada:2018zxn,Hanada:2025rca} and references therein, the confined and deconfined phases can be connected by a partially deconfined (PD) phase, and this phase can be stable or unstable depending on the details of the gauge theory. 
In this context, certain gauge theories possess similar gaps in their phase diagram, which have been observed in the works cited above, and are produced by the effects of partial deconfinement. 

It is interesting to mention that, as described in more detail in \cite{Klebanov:2007ws}, the entanglement entropy as a probe for confinement has some privileges over other methods to study the confinement/deconfinement phase transition. The RT surface under study is only a probe at the zero-temperature background, but studying the thermal phase transitions normally needs finding
a gravity solution that possesses a horizon, calculating its action, and comparing it with a regular solution. Finding these solutions is much more complicated than solving for probes. The conventional RT probe normally divides the space into two regions and can tell us about phase transitions happening in space. In our case, we are able to divide the space of internal degrees of freedom into two regions, hence it can be that we are able to study different phases in the internal color/flavor space. Any phase transition in the $S_\text{EE}$ value might be related to the partial deconfinement of certain degrees of freedom in the quiver. 

One can get more evidence for this conjecture in our study case by paying attention to the bottom-left panel of figure \ref{fig:rxz_solution_multi}, where the profile of the probe is approaching the end of space ($r\to r_*$) only for some values of the $z$ coordinate, but not the whole range. Hence, for certain energies, some degrees of freedom are deconfined, while some are still in the confined/screened structure, and the configuration is shown to be interpolating between the two.

To the best of our knowledge, this calculation and the following observation of the appearance of a gap in the EE of quiver QFTs had not been made before and might open the way for further research.

\FloatBarrier
\section{Conclusions and outlook}\label{concl}

In this work, we computed and analyzed the EE for a family of holographic quiver theories, making use of the RT conjecture. Our results, presented in section \ref{sec:discussion}, show for the first time the calculation of the EE where one of the directions under study is an internal direction parameterizing internal degrees of freedom, in our case $z$, which explores the quiver length on the QFT side of the holographic duality. Our numerical resolution, whose outcome is visible, e.g. in figures \ref{fig:S_Lzs_multiQ} and \ref{fig:Sofzs_multiQ_multiL} for three different quivers, clearly shows novel features, such as a phase transition, the dependency on the quiver direction, the interplay between spatial and quiver directions, and the dependency on the specific quiver.

We now present a brief conclusion and outline potential avenues for future research, building upon the key findings summarized below.

\begin{itemize}
 \item{We provided a review of an infinite family of massive type-IIA backgrounds dual to a family of 4D ${\cal N}=1$ SCFTs. After a deformation and a twisted compactification on a circle, these 4D SCFTs flow to gapped $(2+1)$-dimensional theories with four supercharges. This holographic RG flow is realized on the gravity side by a deformation of the geometry that terminates smoothly.}
 \item{We calculated the entanglement entropy in the field theory side using holographic methods, specifically by optimizing the action of a codimension-two submanifold.} 
 \item{Investigating the EE calculation, we found an interesting dependence of the entropy on the quiver gauge/flavor degrees of freedom.}
 \item{In addition to the conventional calculation of the EE, done by dividing the space direction of the QFT into two subregions, we also divided the internal degrees of freedom along the quiver; this was only made possible by using the powerful and simple geometrical description in hand from the holographic dual background. The results obtained in section~\ref{sec:rofz} are in accordance with the intuitive expectations of the physical theory.}
\item{The method used to obtain the minimal RT surface and the entropy $S_\text{EE}$ as functions of the separations $(L,z_\star)$ was a numerical constrained optimization algorithm to solve the variational problem, and we used splines on triangulations to model the surface by discretization, as described in detail in section~\ref{sec-numerical}. A major achievement, compared to the method used in ref.~\cite{Giliberti:2024eii}, is solving the dynamics for a more involved embedding, depending on two variables and thus yielding a double integration.
}
\end{itemize}

There are interesting lines of research for further consideration.
\begin{itemize}
 \item{A very rich phenomenological interpretation is hidden in the functions $S_\text{EE}(z_\star)$ and $S_\text{EE}(L,z_\star)$, provided in sections \ref{sec:rofz} and \ref{sec:genericem}. It would be desirable to understand, at least to some extent, the pure field-theoretic interpretation of these observations. Of course, the QFT side is strongly coupled, hence making a direct calculation very complicated, which is the reason for our choice of strategy.}
  \item{The phase transitions, or jumps, observed in the 3D $S_\text{EE}(L,z_\star)$ plot of figure \ref{fig:S_Lzs_multiQ} are of particular interest and need to be studied more carefully. Very rich phenomenology, related to the quiver structure of the QFT, is expected to be involved in the observation. Specifically, the partial deconfinement scenario might be of particular interest.}
 \item{A promising direction for future work is the application of our variational formalism to a broader class of non-local observables, including 't Hooft loops and holographic complexity. Since the fundamental actions for these quantities share the generic structure of \eq{AFPRT_metric}, our framework provides a natural starting point. Indeed, this work is already a considerable improvement compared to ref.~\cite{Giliberti:2024eii}, for here we present the ability to describe more complex probe objects.
}
\item{It would be interesting to solve the Euler-Lagrange equations derivable from \eq{action_full} directly, using the numerical or combined analytical methods, to verify our results.}
\item{Similar calculations as the ones presented here can be applied to other families of backgrounds dual to QFTs presented in \cite{Macpherson:2024frt,Akhond:2022awd,Akhond:2022oaf,Legramandi:2021aqv,Akhond:2021ffz,Lozano:2020bxo,Lozano:2020txg}. It could be beneficial to understand how the techniques developed for this work can be implemented in the backgrounds dual to flows between CFTs of different dimensions.}
\end{itemize}

\vspace{0.5cm}

\hrule

\appendix


\section*{Acknowledgments}

For discussions, comments on the manuscript, and for sharing their ideas with us, we wish to thank: Carlos Nunez, Federico Castellani, Aldo Lorenzo Cotrone, Bruno Degli Esposti, and Masanori Hanada. We are especially grateful to Carlos Nunez for suggesting this problem to us.
We thank the University of Florence and Swansea University for the \textsc{Matlab} license used for this work. 
The work of M.G. was funded by the European Union - Next Generation EU - National Recovery and Resilience Plan (NRRP) - M4C2 CN1 Spoke2 - Research Programme CN00000013 ``National Centre for HPC, Big Data and Quantum Computing'' - CUP B83C22002830001. The work of D.C. and M.H. has been supported by the STFC consolidated grant ST/Y509644-1. M.H. would like to thank the Galileo Galilei Institute for Theoretical Physics for the hospitality and the INFN for partial support during which we had conversations about this work. 

The plots for this article were made using the \textsc{Matlab} library \texttt{ProfessionalPlots} \cite{professionalplots}.

\vspace{0.5cm}

{\bf Open Access Statement} --- For the purpose of open access, the authors have applied a Creative Commons Attribution (CC BY) license to any Author Accepted Manuscript version arising. 

\vspace{0.5cm}

{\bf Research Data Access Statement} --- The data generated for this manuscript can be downloaded from \cite{giliberti_2025_17160969}, and the solutions can be visualized in \href{https://www.mathworks.com/matlabcentral/fileexchange/181994-holographic-entanglement-entropy-quiver-solutions}{\texttt{Matlab file exchange}}. A \textsc{Matlab} license is needed to visualize the solutions.

\vspace{0.5cm}

\hrule

\bibliographystyle{JHEP}
\bibliography{main.bib}

\end{document}